\documentclass[12pt,preprint]{aastex}
\shorttitle{Photon Bubbles and Disk Structure}
\shortauthors{Begelman}
\begin{document}
 
\title{Photon Bubbles and the Vertical Structure of Accretion Disks}
\author{Mitchell C. Begelman\altaffilmark{1}}
\affil{Joint Institute for Laboratory Astrophysics, University of Colorado at Boulder}
\affil{JILA, 440 UCB, Boulder, CO 80309-0440} 
\email{mitch@jila.colorado.edu}
\altaffiltext{1}{Also at Department of Astrophysical and Planetary Sciences, University of Colorado}
       
%\date{Accepted .     Received  }
%\pubyear{1999}
 
\begin{abstract}
We consider the effects of ``photon bubble" shock trains on the vertical structure of radiation pressure-dominated accretion disks.  These density inhomogeneities are expected to develop spontaneously in radiation-dominated accretion disks where magnetic pressure exceeds gas pressure, even in the presence of magnetorotational instability (MRI).  They increase the rate at which radiation escapes from the disk, and may allow disks to exceed the Eddington limit by a substantial factor without blowing themselves apart.  To refine our earlier analysis of photon bubble transport in accretion disks, we generalize the theory of photon bubbles to include the effects of finite optical depths and radiation damping.  Modifications to the diffusion law at low $\tau$ tend to ``fill in" the low-density regions of photon bubbles, while radiation damping inhibits the formation of photon bubbles at large radii, small accretion rates, and small heights above the equatorial plane.  Accretion disks dominated by photon bubble transport may reach luminosities of $10->100$ times the Eddington limit ($L_E$), depending on the mass of the central object, while remaining geometrically thin.  However, photon bubble-dominated disks with $\alpha-$viscosity are subject to the same thermal and viscous instabilities that plague standard radiation pressure-dominated disks, suggesting that they may be intrinsically unsteady.  Photon bubbles can lead to a ``core-halo" vertical disk structure.  In super-Eddington disks the halo forms the base of a wind, which carries away substantial energy and mass, but not enough to prevent the luminosity from exceeding $L_E$.  Photon bubble-dominated disks may have smaller color corrections than standard accretion disks of the same luminosity.  They remain viable contenders for some ultraluminous X-ray sources and may play a role in the rapid growth of supermassive black holes at high redshift.    

\end{abstract}


\section{Introduction}

Photon bubbles can greatly enhance the escape of radiation from  magnetized, radiation pressure-dominated atmospheres.  In photon bubble instability (Arons 1992; Gammie 1998), the atmosphere spontaneously forms a propagating pattern of low-density channels separated by regions of high density.  The fastest growing inhomogeneities evolve to nonlinear amplitude, becoming a train of isothermal, gas pressure-dominated shocks (Begelman 2001, hereafter B01; Turner et al. 2005).  Radiation tends to escape through the underdense regions between the shocks, avoiding the regions of high density. The mean radiation flux exceeds the local Eddington limit (calculated using the local gravity, $g(z)$: Shaviv 1998) by a factor that can approach the ratio of magnetic pressure to mean gas pressure, yet overall hydrostatic equilibrium is maintained by magnetic tension and the dynamic cycling of gas between high- and low-density phases.  Because the shocks tend to evolve toward larger separations and higher density contrasts, atmospheres susceptible to photon bubble instability can reach very high luminosities. 

Begelman (2002, hereafter B02) argued that photon bubble shock trains may dominate the escape of radiation from the inner regions of radiation pressure-dominated accretion disks.  Such porous disks would be thinner than standard disks of the same luminosity.  Under extreme conditions, it may be possible for the total luminosity of the disk to exceed the global Eddington luminosity by a large factor, without the disk being destroyed by mass loss. 

In this paper we attempt to flesh out the effects of photon bubbles on the vertical structure of accretion disks, and particularly on the structure of the disks' surface layers and on mass loss in the super-Eddington limit.   Our analysis is based on analytic and numerical calculations of photon bubbles assuming a static background with a uniform magnetic field. Further simulations are required before we can answer the crucial question of whether photon bubbles form and grow in the presence of magnetorotational instability (MRI), which generates a magnetic field that fluctuates in time and space.  Arguments based on instability growth rates (reviewed in \S~3) suggest that the answer is yes.  Other issues are susceptible to analytic treatment now, and we focus on these. The B02 analysis paid too little attention to the effects of finite optical depth and radiation damping on the existence and properties of photon bubbles in accretion disks.  These effects prove to be crucial.  The basic mechanism of photon bubbles relies on the inverse relationship between radiation flux and density, predicted by the diffusion law.  When the optical depth across a region of changing density becomes small, the coupling between flux and local density declines.  This means that the denser regions see a higher flux than predicted by standard diffusion, while the neighboring tenuous regions see a smaller flux. Radiation damping (commonly known as ``Silk damping": Silk 1968) is a separate effect, and results from radiative diffusion driven by the divergence of the velocity field. It is most important at high optical depths.

The effects of finite optical depth become important when the overall optical depth through the atmosphere is still very large, because photon bubbles involve length scales much smaller than the scale height of the atmosphere.  In \S~2 we generalize the nonlinear theory of photon bubbles to include the effects of finite optical depth.  To do this we require a generalization of the radiation diffusion law, to take account of the nonlocal relationship between flux and density at low $\tau$.  The closure we describe in Appendix A is based on the model by Ruszkowski \& Begelman (2003), modified to include radiation damping. We find that the crucial parameter is the optical depth for gas at the mean density, measured across a distance equal to the {\it gas} pressure scale height.  This critical optical depth --- $\tau_0$, in our notation --- is smaller than the total optical depth by a factor of order the ratio of gas pressure to radiation pressure.  Radiation damping impedes the existence of photon bubbles where the radiation diffusion time through the disk is longer than the gas sound crossing time. The ratio of these timescales gives rise to a second parameter, the damping parameter $\xi$ (B01).  (Note that B01 refers to $\xi$ as the ``radiation drag" parameter.  ``Radiation damping" is a more appropriate term than ``radiation drag" in this context, because the underlying radiation field is not fixed but rather responds to the flow; see additional comments following eq.~[\ref{radentropy}].)  Photon bubbles can exist only when $\xi$ is sufficiently small; the relationship between $\xi$ and $\tau_0$ determines the photon bubble behavior, as discussed in \S~2.2.      

We apply the generalized theory of photon bubbles to radiation pressure-dominated accretion disks in \S~3.  Below a certain accretion rate, corresponding to a luminosity of a few percent to tens of percent of the Eddington limit (depending on $\alpha$ and the mass of the central object), radiation damping is too strong for photon bubbles to dominate the disk. At higher accretion rates, photon bubbles begin to form within the radius where $\xi \sim 1$.  Since the radiative losses associated with photon bubbles are expected to increase strongly with decreasing $\xi$ at $\xi \la 1$, we argue that there should be a feedback mechanism that keeps the disk core at $\xi \sim 1$ at small $r$.  However, we also find that this self-regulated photon bubble loss mechanism is subject to the same thermal and viscous instabilities that plague standard radiation pressure-dominated $\alpha-$disks.  We reaffirm our earlier result that steady state disks can exceed the Eddington limit while remaining thin.  

In \S~3.2 we show that disks dominated by photon bubble transport should form a vertical ``core--halo" structure.  The halo has a much lower density and pressure than the core, yet possesses enough optical depth to maintain photon bubbles well outside the core.  Where the luminosity is sub-Eddington, the core smoothly merges into a homogeneous atmosphere at the expected height above the equator.   For super-Eddington luminosities the halo forms the base of a wind (\S~3.3).  We argue that the mass loss, while substantial, is unlikely to prevent highly super-Eddington accretion.  In effect, the halo insulates the disk core against excessive mass loss.  In \S~3.4, we comment on the effects of photon bubble transport on the color temperatures of disks. We summarize our results in \S~4, where we also discuss important caveats and amplify on two potential applications of the model: to ultraluminous X-ray sources and the rapid growth of supermassive black holes.
 
\section{Photon Bubbles at Finite Optical Depth}

The essence of the photon bubble mechanism is the inverse relationship between radiation flux and density. At large optical depths the diffusion law for flux is  ${\bf F} = - (c/\kappa) \nabla p / \rho $, where $\kappa$ is the opacity (assumed constant) and $p$ is the radiation pressure.  In order to maintain hydrodynamical equilibrium, pressure gradients must be relatively weak, but density contrasts can be very strong. Thus, radiation flux is excluded from dense regions while passing readily through regions of low density.  In a photon bubble wave train, the densest regions receive so little flux that they are supported against gravity by gas pressure gradients, whereas the low-density regions see a super-Eddington flux and accelerate rapidly.   

This picture breaks down where large changes of density occur across regions of low optical depth, i.e., where $\rho^2 \kappa / |\nabla \rho| <1$.  The flux cannot adjust over a length scale smaller than the photon mean free path.  Where density gradients are large, the dense regions will receive more flux than predicted by the usual diffusion law while the tenuous regions will receive less.  We stress that we are considering localized density gradients within an otherwise optically thick atmosphere.  Thus, to a good approximation the radiation field is everywhere isotropic.  It is the {\it slight} anisotropy, giving rise to the flux, that becomes decoupled from local changes in the density. 

In this section we consider how the structure of a photon bubble shock train changes as the optical depth across the bubble decreases.  Effects of finite optical depth set in while the total optical depth across the bubble --- $\tau_\lambda = \rho_0 \kappa \lambda$, where $\rho_0$ is the mean density and $\lambda$ is the spacing between shocks --- is still large.  We shall see below that the critical optical depth, $\tau_0$, is the much smaller optical depth across a gas pressure scale height, $\tau_0 = \rho_0\kappa c_g^2/ g$, where $c_g^2$ is the gas thermal speed and $g$ is the local gravitational acceleration.  For $\tau_0 >1$, photon bubbles are essentially unaffected by optical depth effects.  For $\tau_0 < 1 < \tau_\lambda$, the density contrasts within photon bubbles are smaller, but the basic mechanism persists.  The weakening continues until $\tau_\lambda \la 1$, at which point the photon bubble mechanism fails.   

The remainder of \S~2 is more technical than later parts of the paper.  Readers who are less interested in the physics of photon bubbles may wish to note the definitions in \S~2.1.1 and then skip to \S~2.2.3.  The principal results needed for the accretion disk discussion in \S~3 are found in Table 1 and \S~2.2.4.   

\subsection{Equations}

Our analysis is based on a simplified set of radiation-hydrodynamical equations, in which the magnetic field is stiff and is inclined to the $x-$axis by an angle $\theta$, where $-\pi/2 < \theta < \pi/2$. We assume that all vectors lie in the $x-z$ plane, and that there is a uniform gravitational field $- g \hat z$. Since the motion is constrained to lie along ${\bf B}$, the components of the velocity are ${\bf v} = v(\cos\theta, \sin\theta)$.  As in B01, we separate the gas and radiation energetics and assume an isothermal equation of state for the gas, with a constant sound speed $c_g = (p_g / \rho)^{1/2}$.  For additional discussion of the approximations and assumptions behind this model, see B01.

The five equations that must be satisfied describe continuity, momentum conservation parallel to ${\bf B}$, conservation of radiation energy, and radiative transfer in both the $x$ and $z$ directions.  We seek periodic ``plane-wave" solutions with wavelength $\lambda$, in which quantities associated with the wave depend on position and time through the combination
\begin{equation}
\label{sdef}
s = x \cos\phi + z \sin\phi  + v_p t ; 
\end{equation}
until further notice we denote differentiation with respect to $s$ by a prime. We may arbitrarily assume $v_p > 0$, i.e., that wavefronts move from right to left.  Thus, the wavevector makes an angle $\phi + \pi$ with respect to the $x-$axis, where $-\pi/2 < \phi < \pi/2$. The quantity $v_p$ is the phase speed of the wave; note that the wave speed used in B01, $v_0 \equiv v_p / \cos\phi$, is the speed of the intersection of a wave front with the $x-$axis. For purposes of comparison, note also the following replacements of the parameters $\zeta$ and $b$ from B01: $\zeta = \tan\phi$, $b = \tan\theta$.  According to our conventions, $\cos\phi$ and $\cos\theta$ are always positive, while $\sin\phi$ and $\sin\theta$ may take on either sign.

We consider photon bubbles with wavelengths much smaller than the radiation pressure and density scale heights, in which buoyancy forces are unimportant.  (See Arons [1992] and Begelman [2006] for discussions of photon bubbles driven by buoyancy.) To the required level of approximation we may treat the velocity and density as pure functions of $s$. Defining $z$-independent wave-averaged quantities by the integral 
\begin{equation}
\label{meandens}
\langle A \rangle \equiv \lambda^{-1} \int^{s+\lambda}_s A(s)  ds ,
\end{equation}
we define the mean density $\langle \rho \rangle = \rho_0 $. In order to maintain overall hydrostatic equilibrium of the atmosphere, we cannot ignore the secular variation of pressure with height.  For the radiation pressure we write $p(z, s) = p(s) - \rho_0 g z$,  with $\langle p \rangle = p_0$.  The ``isothermal" sound speed associated with the radiation-dominated atmosphere is $c_0 = (p_0/\rho_0)^{1/2}$ and  $H = c_0^2 / g $ is the radiation pressure scale height of the atmosphere. 
 
\smallskip
\noindent \underbar{\it Continuity}. The continuity equation, 
\begin{equation}
\label{}
{ D\rho \over Dt} = - \rho \nabla\cdot \bf v ,
\end{equation}
yields, to lowest order,  
\begin{equation}
\label{cont}
v_p\rho' + \cos(\theta - \phi)(\rho v)'  = 0  ,
\end{equation}
which is readily integrated. To determine the constant of integration, we 
demand that there be no net mass flux through the atmosphere, $\langle \rho v \rangle = 0$. 
We then have
\begin{equation}
\label{cont2}
\rho \left[v_p + \cos(\theta - \phi) v \right] =  \rho_0 v_p.
\end{equation} 

\smallskip
\noindent \underbar{\it Momentum parallel to ${\bf B}$}. Projecting the momentum equation onto the magnetic field direction, we obtain
\begin{equation}
\label{mom1}
\rho v' \left[v_p + \cos(\theta - \phi) v \right] = \rho_0 v_p v' = - c_g^2 \rho' \cos(\theta - \phi)  - \rho g  \sin\theta + {\rho \kappa \over c} \left( F_x \cos\theta + F_z \sin\theta \right).
\end{equation} 
 
\smallskip
\noindent \underbar{\it Energy}. The radiation energy equation can be written
\begin{equation}
\label{radentropy}
\nabla \cdot {\bf F} = - 4 p \nabla\cdot {\bf v} - 3 {Dp\over Dt} . 
\end{equation}
Under conditions likely to apply in accretion disks we may ignore the last term on the right-hand side. The first term on the right-hand side leads to damping via radiative diffusion (``Silk damping" [Silk 1968]; described as the ``nondiffusive regime" in section 2.1 of Agol \& Krolik 1998; see also Blaes \& Socrates 2003 for an extensive discussion in relation to linear instabilities of radiation-dominated atmospheres).

Assuming that $\bf F$ is a function of $s$, we have  
\begin{equation}
\label{divF}
\nabla \cdot {\bf F} + 4 p \nabla \cdot {\bf v} = (F_x + 4 p v \cos\theta)' \cos\phi + (F_z +4 pv \sin\theta)' \sin\phi = 0.  
\end{equation}
In order to maintain wave-averaged hydrostatic equilibrium in the momentum equation, eq.~(\ref{divF}) must integrate to 
\begin{equation}
\label{Fint}
F_z \sin \phi + F_x \cos\phi +  4 pv \cos (\theta - \phi) = {gc \over \kappa} \sin\phi = F_E \sin\phi,
\end{equation}
where $F_E = gc/\kappa$ is the Eddington flux. 

\smallskip
\noindent \underbar{\it Radiative transfer}. Because we wish to consider the effects of low optical depths within photon bubbles, the usual diffusion equation is not applicable. Instead, we will use the closure developed by Ruszkowski \& Begelman (2003; hereafter, RB), as generalized in Appendix A to handle motion.  Tests of the stationary (RB) version against Monte Carlo simulations suggest that this closure adequately captures the main features of radiative transfer in the presence of tilted, plane-parallel density variations deep within an atmosphere.  Using equations (\ref{divTmoment2}) and (\ref{Tij}), we obtain
\begin{equation}
\label{Fx}
F_x = - {c\over \rho \kappa} {\partial p \over \partial x} + {1 \over 5 \rho\kappa} \left[ 2 {\partial \over \partial x} \left( {1\over \rho\kappa} {\partial F_x \over \partial x}\right) + {\partial \over \partial z} \left( {1\over \rho\kappa} {\partial F_z \over \partial x} + {1\over \rho\kappa} {\partial F_x \over \partial z}\right) \right] ,
\end{equation}
with an analogous equation for $F_z$.  Using eq.~(\ref{divF}) we can show that  
\begin{equation}
\label{Fx2}
F_x = - {c\over \rho \kappa} {\partial p \over \partial x} + {1 \over 5 \rho\kappa} \left[\left( {F_x' \over \rho\kappa }\right)' - 4\cos(\theta-\phi) \cos\phi \left( {p v' \over \rho\kappa }\right)' \right]
\end{equation}
\begin{equation}
\label{Fz2}
F_z = - {c\over \rho \kappa} {\partial p \over \partial z} + {1 \over 5 \rho\kappa} \left[ \left( {F_z' \over \rho\kappa }\right)' 
- 4\cos(\theta-\phi) \sin\phi \left( {p v' \over \rho\kappa }\right)' \right]
\end{equation}
and by using eq.~(\ref{Fint}) and the adopted form of $p(z,s)$ we have
\begin{equation}
\label{pprime}
p' = \left(g \sin\phi + {4 p_0 \kappa v_p \over c} \right)(\rho_0 - \rho)  .
\end{equation}
  
\subsubsection{Dimensionless Equations}

At this point, it makes sense to simplify the notation by expressing the equations in terms of dimensionless variables.  We define: 
\begin{equation}
\label{nondim}
\eta \equiv {\rho \over \rho_0}; \ \  f  \equiv {\kappa F_x\over g c \sin\phi} ; \ \  w \equiv {s g \over c_g^2} \equiv {s \over H_g};  
\ \ m_p \equiv {v_p\over c_g | \cos(\theta - \phi)|} ; \ \ \tau_0 \equiv \rho_0 \kappa H_g  ; \nonumber
\end{equation}
\begin{equation}
\label{morenondim}
\beta \equiv {c_g^2 \over c_0^2}; \ \  M_0  \equiv {c \over \kappa \rho_0 H c_0  } ; \ \  \xi \equiv { 4 \beta^{1/2} | \cos(\theta - \phi)| \over M_0 } . 
\end{equation}
These scalings are physically motivated.  We normalize $s$ to the gas pressure scale height $H_g = c_g^2/g$, which is the characteristic scale in the high-density region of the photon bubble.  The quantity $m_p$ is a Mach number expressing the speed at which projected wave fronts move along a magnetic field line, normalized to the gas pressure sound speed.  $\tau_0$ is the optical depth across a distance $H_g$, assuming that the gas has its mean density $\rho_0$.  $M_0$ is the diffusivity parameter from Arons (1992), roughly the ratio of the dynamical time to the diffusion time across the atmosphere, $c/\tau_H c_0$, where $\tau_H$ is the optical depth across the radiation pressure scale height. The damping parameter $\xi \sim \tau_Hc_g/c$ is similar to the quantity referred to as the ``drag parameter" and denoted by the same symbol in B01; it physically represents the ratio of diffusion time to sound crossing time in the gas, measured across the radiation pressure scale height.  All dimensionless quantities except $f$ are positive-definite.
 
In dimensionless form, the radiative transfer equation (\ref{Fx2}) becomes 
\begin{equation}
\label{nondim2}
f = \cos \phi \left( 1 - {1 \over \eta} \right) \left( 1 + {\xi m_p \over \sin\phi}\right)  + {1\over 5 \tau_0^2 \eta} \left[ \left( {f' \over \eta} \right)' + \xi m_p \cot\phi  \left( {\eta' \over \eta^3} \right)' \right] ,
\end{equation}
where a prime now denotes differentiation with respect to $w$.  Eliminating $F_z$ in favor of $F_x$ and using eq.~(\ref{nondim2}) to eliminate $f$ in favor of its derivatives,  the momentum equation (\ref{mom1}) becomes 
\begin{equation}
\label{nondim3}
\left( {m_p^2\over \eta^2} - 1 \right) {\eta' \over \eta} = \left( 1 - {1\over \eta} \right)\left[\cos\phi \tan (\theta - \phi) - \xi m_p \right]  + {\tan(\theta-\phi)\over 5 \tau_0^2 \eta} \left[ \left( {f' \over \eta} \right)' + \xi m_p \cot\phi  \left( {\eta' \over \eta^3} \right)' \right] .
\end{equation} 
Equations (\ref{nondim2}) and (\ref{nondim3}) describe photon bubbles with finite optical depth and radiation damping. 

\subsection{Analysis}

Equations (\ref{nondim2}) and (\ref{nondim3}) can be combined into a third-order, nonlinear ordinary differential equation for $\eta (w)$.  Physical solutions must pass through a critical point at $\eta^2 = m_p^2$ and a shock transition with the jump condition $\eta_+ \eta_- = m_p^2$, where $\eta_+$ and $\eta_-$ are the post-shock and pre-shock densities, respectively.  The quantities $f$ and $f'/\eta + \xi m_p \cot\phi (\eta'/\eta^3)$ must be continuous across the shock front. 

We do not attempt a full numerical solution of equations (\ref{nondim2}) and (\ref{nondim3}).  Instead, we use approximate solutions to estimate the quantitative changes in photon bubble structure as $\tau_0$ decreases. We restrict our attention to photon bubbles with large density contrasts, $\eta_- \ll 1 \ll \eta_+$.

\subsubsection{Infinite Optical Depth}

In the limit $\tau_0 \rightarrow \infty$, eq.~(\ref{nondim3}) reduces to a single first-order equation in $\eta$, 
\begin{equation}
\label{nondim5}
\left( {m_p^2\over \eta^2} - 1 \right) \eta'  =  \left( \eta - 1 \right) \left[ \cos\phi \tan (\theta - \phi)  - \xi \right],
\end{equation} 
recovering the ``stiff-wire" limit discussed in B01 and Turner et al.~(2005).  

The critical condition and jump condition imply that 
\begin{equation}    
\label{jump}
m_p^2 = \eta_+ \eta_- = 1.
\end{equation}
Between the shock and the critical point (where $\rho = \rho_0$), the gas is supported against the effective gravity primarily by gas pressure and the density profile is approximately exponential, $\rho \approx \rho_+ \exp\{- w [\cos\phi\tan (\theta - \phi)-\xi]\}$.  Beyond the critical point, the density varies inversely with distance, $\eta \approx \{w [\cos\phi\tan (\theta - \phi)-\xi ]\}^{-1}$.  The characteristic optical depth in the low-density region, $\tau (w) = \tau_0 \eta w = \tau_0 [\cos\phi\tan (\theta - \phi)-\xi]^{-1}$, is constant.  If $\tau_0 \gg 1$, the optical depth is large everywhere and we are justified in ignoring finite-$\tau$ effects.

\subsubsection{Onset of Finite-$\tau$ Effects}

The effects of finite optical depth appear first near the critical point and in the low-density region downstream.  Before tackling the case where these effects are dominant, let us first consider them in the perturbative limit. To simplify the algebra, in this section we consider only the case $\xi = 0$.  Let
\begin{equation}
\label{etazero}
\eta = \eta_0 (1 + \varepsilon \eta_1) , \ \ f = f_0 (1 + \varepsilon f_1), \ \  m_p^2 = 1 + 2 \varepsilon m_1, 
\end{equation}
where $\varepsilon \equiv (5 \tau_0^2)^{-1} \ll 1$ and $\eta_0$, $f_0$ are the solutions of equations (\ref{nondim2}) and (\ref{nondim3}) in the limit $\varepsilon \rightarrow 0$, i.e., 
\begin{equation}
\label{etazero2}
{\eta_0' \over \eta_0^2 } (1 + \eta_0) = -\cos\phi\tan (\theta - \phi), \ \ f_0  = {\eta_0 - 1 \over \eta_0} \cos\phi  .
\end{equation}
Note that $\eta_1$ and $f_1$ are functions of $w$ while $m_1$ is a constant.  Expanding equations (\ref{nondim2}) and (\ref{nondim3}) to $O(\varepsilon)$, we obtain
\begin{equation}
\label{etaone}
\left( {1 \over \eta_0^2} - 1 \right) \left( \eta_0 \eta_1' + \eta_0' \eta_1 \right) + {2(m_1 - \eta_1)\eta_0' \over \eta_0^2 } = \left[ \eta_0 \eta_1 \cos\phi + \left( {f_0' \over \eta_0}\right)'\right] \tan (\theta - \phi)  
\end{equation}
and
\begin{equation}
\label{fone}
\left[ \eta_0 \eta_1 \cos\phi + \left( {f_0' \over \eta_0}\right)'\right] = \eta_0 f_0 (f_1 + \eta_1)  . 
\end{equation}

Equation (\ref{etaone}) still has a critical point at $\eta_{\rm 0, cr}=1$, $f_{\rm 0, cr}=0$; from eq.~(\ref{etazero2}) we determine that
\begin{equation}
\label{critconds}
\eta'_{\rm 0, cr}=- {1\over 2}\cos\phi \tan(\theta - \phi), \ \ \eta''_{\rm 0, cr}= {3\over 8} \cos^2\phi \tan^2 (\theta - \phi) , \ \ 
 \left( {f_0' \over \eta_0}\right)'_{\rm cr} = - {3\over 8} \cos^3\phi \tan^2 (\theta - \phi) .
\end{equation}
At the critical point the left-hand side of eq.~(\ref{fone}) must vanish, while eq.~(\ref{etaone}) implies $m_1 = \eta_{\rm 1, cr}$.  Therefore, 
\begin{equation}
\label{critconds2}
m_1 = \eta_{\rm 1, cr}= {3\over 8} \cos^2\phi \tan^2 (\theta - \phi) .
\end{equation}
The fact that $m_1 , \ \eta_{\rm 1,cr} > 0$ means that finite-$\tau$ effects increase both the wave speed and the density at the critical point.  These trends occur because extra radiation flux spills over into the dense region, compared to the usual diffusion law, when the optical depth is finite. Consequently, radiative acceleration takes over from gas pressure support at a higher density, pushing the flow through the critical point sooner and propelling the wave.

We next consider the flow well downstream of the critical point.  For $\eta_0$ and $f_0$ we use the asymptotic solution discussed in \S~2.2.1: $\eta_0 = [w \cos\phi\tan (\theta - \phi)]^{-1}$, $f_0 = - w \cos\phi\tan (\theta - \phi)$.  Assuming $\eta_0 \ll 1$, equations (\ref{etaone}) and (\ref{fone}) yield $w \eta_1' + \eta_1 + {1\over 4} \cos^2\phi\tan^2(\theta - \phi)=0$.  Given that $\eta_1$ is pinned to $\eta_{\rm 1, cr}$ at $w \sim$ a few, the solution at $w \gg 1$ must converge to a constant value, 
\begin{equation}
\label{etaoneasymp}
\eta_1 = - {1\over 4}\cos^2\phi \tan^2 (\theta - \phi) .
\end{equation}
Thus, at large but finite $\tau_0$, the density is depressed below its infinite-$\tau$ value at $w \gg 1$.  However, we will see below that this behavior reverses as $\tau_0$ decreases to the point where the finite-$\tau$ corrections are no longer perturbative.  At small $\tau_0$, $\eta \propto (\tau_0 w)^{-1}$ becomes very much larger than its infinite-$\tau$ value.

Although the density at $w \gg 1$ is decreased by finite-$\tau$ effects, the minimum density of the photon bubble, $\eta_-$, is higher, given a fixed maximum density $\eta_+$. The shock jump condition $\eta_- = m_p^2/ \eta_+$ implies that $\eta_-$ increases with $m_p^2$.  Thus, finite-$\tau$ effects tend to wash out the density contrasts of photon bubbles, a trend we shall continue to find as $\tau_0$ becomes small.      

\subsubsection{Arbitrary Optical Depth}

Guided by our results above, which show that the $\eta \propto w^{-1}$ scaling is preserved to lowest order when finite-$\tau$ effects are included, we seek a solution of the form $\eta \approx A/w$ well downstream of the critical point, i.e., when $\eta \ll m_p$.  In this limit, both $(f'/\eta)'$ and $(\eta'/ \eta^3)'$ are constant.  To model solutions that pass through the critical point, we extrapolate the assumed $A/w$ dependence to the critical pont at $\eta = m_p$, and demand that the right-hand side of eq.~(\ref{nondim3}) vanish at this point.  This implies that
\begin{equation}
\label{nondim4condition}
1 - {\tan(\theta-\phi) \over 5 \tau_0^2 [\cos\phi\tan(\theta-\phi) -\xi m_p]} \left[ \left( {f' \over \eta} \right)' + \xi m_p \cot\phi  \left( {\eta' \over \eta^3} \right)' \right]  = m_p.
\end{equation}   
Extracting the factor $1 - m_p / \eta $ from eq.~(\ref{nondim3}) and returning to the limit $\eta \ll m_p$,  we are left with
\begin{equation}
\label{nondim4condition2}
- m_p{\eta' \over \eta^2} = \cos\phi\tan(\theta-\phi) - \xi m_p .
\end{equation}   

Setting $\eta = A/w$, eq.~(\ref{nondim4condition2}) allows us to express $A$ in terms of $m_p$, 
\begin{equation}
\label{Aequation}
A = {m_p \over \cos\phi\tan(\theta - \phi) - \xi m_p} ,
\end{equation}
while equation (\ref{nondim2}) gives  
\begin{equation}
\label{fequation}
f' =  - {\sin\theta\over \sin\phi\sin(\theta - \phi)}{\xi m_p \over A} - {m_p^2 \over A^2 \tan(\theta - \phi)} .
\end{equation}
Since $A$ must be positive, solutions are possible only for $\cos\phi\tan(\theta - \phi) > \xi m_p > 0$. 
Substituting from eq.~(\ref{fequation}) into eq.~(\ref{nondim4condition}), we obtain
\begin{equation}
\label{mequation}
m_p = 1 + {m_p \over 5 \tau_0^2 A^2} \left\{ 1 + \xi {A \over m_p} \left[ 1 + 2\cot\phi \tan(\theta - \phi) \right]   \right\} .
\end{equation}
Equation (\ref{Aequation}) can then be used to turn this into a quadratic equation for $m_p$.   

If $\xi \ll \min[\cos\phi\tan(\theta - \phi),\ \sqrt{5} \tau_0]$, we may ignore the effects of damping (i.e., set $\xi =0$).  Equation (\ref{mequation}) then has the solution 
\begin{equation}
\label{msolution1}
m_p = {1 \over 2 } \left[ 1 + \left(1 + {4 \cos^2\phi\tan^2(\theta - \phi)\over 5 \tau_0^2 } \right)^{1/2}  \right]   .
\end{equation}
In the perturbative limit $\tau_0 \gg 1$, we obtain a value for $m_1$ greater than eq.~(\ref{critconds2}) by a factor 8/3, but the angular dependence is the same.  The discrepancy in the coefficient reflects the crudeness of our method for approximating solutions that pass through the critical point.

To simplify the analysis further, we specialize to the case $\theta = 0$, which is appropriate for conditions inside accretion disks where the predominant magnetic field is expected to be roughly parallel to the equatorial plane (Balbus \& Hawley 1998).  We then require $\sin\phi < 0$, so that $\tan(\theta - \phi) > 0$. (This restriction implies that photon bubble wavefronts propagate upwards, as seen in simulations [Turner et al. 2005].) Defining $\mu \equiv |\sin\phi|$ and using $\varepsilon \equiv (5\tau_0^2)^{-1}$ from \S~2.2.2 without assuming its magnitude, we obtain the quadratic 
\begin{equation}
\label{mequation2}
m_p^2 (1 - \varepsilon \xi^2) + m_p \left[2 \varepsilon \xi \mu - (1+ \varepsilon \xi^2) \right] - \varepsilon \mu (\mu - \xi ) = 0 .
\end{equation}
Since we identify the critical point with the effective sound speed (which can be boosted above the gas sound speed by radiation pressure), we seek solutions with $1 < m_p < \mu / \xi$, where the upper limit guarantees that $A>0$.  This means that there are no photon bubbles, or very weak ones, for $\xi > \mu$ --- the radiation damping is too strong.  We note that the same restriction applies to the development of the linear photon bubble instability (Gammie 1998; Blaes \& Socrates 2001, 2003). Assuming that $ \mu \sim O(1)$, we may solve eq.~(\ref{mequation2}) approximately for the three characteristic orderings of $\tau_0$, $\xi$, and 1 that admit nonlinear photon bubble solutions according to our simple ansatz.  The properties of these solutions are shown in Table 1.

\begin{deluxetable}{lcccl}
\tablecaption{Parameter Space for Photon Bubbles}
\tablehead{\colhead{Case}&\colhead{Ordering}&\colhead{$m_p$}&\colhead{$A$}&\colhead{Description}}
\startdata
a & $\xi \ll 1 \ll \tau_0$ & $1$ & $1/\mu$ &Ideal, undamped\\
b & $\xi \ll \tau_0 \ll 1$ & $\mu /\sqrt{5} \tau_0$ & $1/\sqrt{5} \tau_0$ & Diffusion-dominated, undamped\\
c & $\tau_0 \ll \xi \ll 1$ & $\mu/\xi$ & $\max\left[1/\sqrt{5}\tau_0,\ \mu/\xi^2 \right]$& Diffusive, critically damped\\  
\enddata
\end{deluxetable}  

Case (a), labeled ``ideal", was treated in B01 and Turner et al.~(2005).  In this solution the standard diffusion approximation is valid everywhere.  The other cases involve modifications to the diffusion law, due to low optical depths.  Case (b) illustrates the effects of low $\tau_0$ in the limit where radiation damping is unimportant. In this case $m_p  \gg 1$, continuing the trend toward larger wave speeds that we saw in \S~2.2.2.  A higher wave speed, in turn, implies that the density contrast is smaller, from the shock jump condition $\eta_+\eta_- = m_p^2$ with $\langle\eta\rangle = 1$. Also, note that $A \approx (\sqrt{5} \tau_0)^{-1} \gg 1$, reversing the trend toward lower density that we found in the perturbative limit.   The effects of low optical depth cause the low-density region of the photon bubble to ``fill in", at least in the undamped limit.  Note that the characteristic optical depth across a density scale length is $A\tau_0 \sim 5^{-1/2}$.  Thus, the optical depth across the low-density part of the wave pattern is regulated to be $\sim 1$.  Case (c) represents another situation in which the wave speed can be much larger than the gas sound speed.  In this case, radiation damping approximately balances the accelerating forces, hence the label ``critically damped".  $A$ is large, and the bubble is once again filled in --- here the characteristic optical depth is at least $\sim 1$ and may be much larger.  In the limit $\xi \rightarrow \mu$, $m_p$ approaches 1 but the density remains large provided that $\tau_0 \ll \xi$.  
 
We can use the expressions for $m_p$ and $A$ to estimate the wavelength of the photon bubble as a function of the density on either side of the shock front.  Assuming that we may extrapolate the density profile $\eta \approx A/w$ all the way to the shock, we have $\lambda \approx A/\eta_-$.  The shock jump condition then gives 
\begin{equation}
\label{lambdaestimate}
\lambda \approx {A\over m_p^2} \eta_+ .
\end{equation}
It will prove convenient to express $\lambda$ in terms of the maximum density $\eta_+$.  Simulations suggest that nonlinear photon bubbles tend to evolve toward the longest permitted wavelength (Turner et al.~2005), corresponding to the largest permitted value of $\eta_+$. The latter is either set by the magnetic field strength, which must be large enough to resist buckling under the postshock gas pressure, or the wavelength, which must be smaller than the scale height.   
 
\subsubsection{Mean Flux}

For the purpose of studying the release of radiation by accretion disks, we are interested in the vertical flux as measured in the laboratory frame, 
\begin{equation}
\label{labflux}
\langle F_{z, {\rm lab}}\rangle = \langle F_z + 4 p v \sin\theta \rangle .
\end{equation}
Since the local flux increases linearly with $w$ in the low density part of the bubble, the mean flux is given approximately by the local flux at $w = \lambda /2$. Since we are assuming $\theta = 0$, the second term on the right-hand side of eq.~(\ref{labflux}) does not contribute.  From equations (\ref{nondim3}) and (\ref{Fz2}), we obtain the Eddington enhancement factor
\begin{equation}
\label{Eddfactor}
\ell \equiv { \langle F_{z, {\rm lab}}\rangle \over F_E } \approx 1 + {1 - \mu^2 \over 2 \eta_-} \left\{ 1  + \xi m_p {\mu\over 1-\mu^2} + {m_p^2\over 5 \tau_0^2 A^3 \mu} \left[1 - \xi {A\over m_p}  \right]  \right\}.
\end{equation}
 
\section{Application to Accretion Disks}

Can nonlinear photon bubbles form in accretion disks, and if so, do they affect the disk structure and luminosity?  The answer depends largely on the relationship between the dissipation rate and the density as a function of height above the disk equator.  Simulations of magnetohydrodynamic disks, both with (Turner, Stone, \& Sano 2002; Turner et al.~2003; Turner 2004) and without (Miller \& Stone 2000; Hirose et al.~2004) radiation transport, suggest that most of the dissipation may occur in a magnetically dominated corona, extending to several density scale heights.  If most of the radiation is produced so high in the corona that the optical depth is low (e.g., Socrates \& Davis 2005, and references therein), then the role of photon bubble transport is moot.  If, however, the bulk of the energy is produced sufficiently close to the equator that $\tau$ is large, then photon bubbles can play a key role in transporting radiation to the disk surface.   The least favorable case for photon bubble transport is the one-zone slab model, in which most of the dissipation occurs in the dense core of the disk.  In this paper we restrict our attention to the latter model. B02 considered this case but did not take into account the effects of radiation damping, which prove to be important at large radii and close to the equator.  In \S~3.1, we will re-analyze the effects of photon bubbles on the structure of disk cores, in the light of our analysis in \S~2.  

Crudely, photon bubble shock trains can be treated as localized phenomena within radiation pressure-supported atmospheres.  Since they propagate at roughly the gas sound speed, they would require many orbital times to cross the disk scale height.  MRI rearranges the magnetic field on a timescale of order the orbital time, suggesting that photon bubble wave trains are continuously being destroyed and reformed before they have time to traverse great distances.  On the other hand, photon bubbles typically grow to nonlinear amplitudes in less than an orbital time (see below).  Simulations show that short wavelength modes quickly evolve toward longer wavelengths and larger density contrasts, until growth is quenched by either buckling of the magnetic field or lack of space (i.e., when the wavelength exceeds the scale height) (Turner et al.~2005).  Where shock fronts extend far enough in height to sample a range of disk conditions, they curve in order to track the local wavevector of maximum instability.  These properties allow us to consider the effects of photon bubbles on disks as a function of height $z$ above the equatorial plane.  In \S~3.2 we will consider the gradual disappearance of photon bubbles with height, as the atmosphere becomes too tenuous to support large local variations in radiation flux.  Even if the core supports large Eddington factors $\ell \gg 1$ due to the presence of photon bubbles, above some height the disk must behave like a homogeneous disk --- material is levitated until the local flux equals the Eddington flux in the effective gravitational field. The transition from a porous to a smooth atmosphere is especially interesting if the core flux exceeds the Eddington limit: in this case, material from the upper layers of the disk cannot remain in  equilibrium and is blown off in a wind.  In \S~3.3 we will sketch the possible nature of this wind, and the extent to which it is likely to disrupt the accretion flow. 

Because the magnetorotational instability (MRI) leads to turbulence, with magnetic field fluctuations on timescales of order the orbital time, there is uncertainty as to whether photon bubbles can actually grow and saturate in accretion disks.  Several authors have argued that photon bubbles are likely to develop even in the presence of the time-varying field, because the linear growth rates of photon bubble instability are typically faster than MRI growth rates (Gammie 1998; Blaes \& Socrates 2001, 2003; B01; B02; Turner et al.~2005).  To quantify this statement, we consider the maximum photon bubble growth rate using Blaes \& Socrates (2003, eq.~[93]), in the limit of tight thermal coupling, a stiff magnetic field, and electron scattering opacity. In terms of our variables, the imaginary part of the frequency is
\begin{equation}
\label{BS03}
\omega_I = {g |\cos (\theta-\phi)| \over 2 c_g} \left[\cos\phi |\tan(\theta - \phi)| - \xi  \right] .
\end{equation}   
This rate is an asymptotic limit for wavelengths $\lambda < H_g$; at longer wavelengths the growth rate declines $\propto \lambda^{-1/2}$.  For the interior of an accretion disk with negligible damping, the maximum growth rate is roughly $\beta^{-1/2} \Omega / 2 $ (for a vertical field), which is much larger than the maximum MRI growth rate $3 \Omega /4$ in the presence of a vertical field (Balbus \& Hawley 1998).  For a horizontal field the maximum photon bubble growth rate is a factor of 2 smaller. The maximum photon bubble growth rate only becomes comparable to the MRI growth rate for wavelengths approaching the disk scale height.  

It is not clear whether long-wavelength ($\lambda \gg H_g$) photon bubble shock trains --- those associated with large density contrasts --- grow directly from linear instabilities, or from the merging of shorter wavelength shock trains.  Simulations (Turner et al.~2005) seem to indicate that the latter process might dominate, in which case long wavelength shock trains could develop even more quickly than predicted by the extrapolation of linear stability theory.

We note that the numerical growth rates presented in Turner et al. (2005: especially Figures 8, 11, and 13) are normalized to $\Omega$, and give the impression that photon bubble growth rates do not exceed a few times $\Omega$.  For purposes of comparison with MRI, this is misleading, because the results presented correspond to a case with rather strong damping, and therefore strong suppression of the growth rate.  Across the disk surface layer considered by Turner et al. (2005), $\xi$  ($=4|\cos(\theta - \phi)| c_g E/ 3 F$ in the notation of Blaes \& Socrates 2003 and Fig.~3 of Turner et al.~2005) decreases with height from 0.53 to 0.43 for the modes plotted in Fig.~8, which have $\theta = 0$, $\phi = 63^\circ$. This is close to the fastest growing mode under these conditions, and the growth rate is $\sim (1.5-2)\Omega$. If damping were negligible, the fastest growing mode for $\theta = 0$ would have $\phi = 45^\circ$, and the growth rate would be $\sim 5.2 \Omega$.  Thus, damping cuts the growth rates in the Turner et al.~(2005) simulation by a factor $\sim 2-3$. 

In addition to time-dependence, MRI and other instabilities create fluctuations in magnetic field direction and strength. Photon bubbles are insensitive to magnetic field strength provided that the magnetic pressure exceeds the maximum gas pressure in the wave train. But fluctuations in direction will distort and refract photon bubble wavefronts (depending on the fluctuation scale compared to the wavelength), and may lead to large fluctuations in photon bubble growth rate and disk porosity.  Where damping is present, a sudden change in field direction could damp out (or enhance) a wave train.  These effects are beyond the current state of analytical models. Effects due to fluctuations in field direction may be largely mitigated by the fact that MRI leads to a predominantly azimuthal field (e.g., Turner 2004), with relatively modest fluctuations in other directions.  Moreover, a strong azimuthal field can slow (Blaes \& Socrates 2001) or even quench (Pessah \& Psaltis 2005) the growth of MRI modes.  
 
In the absence of further guidance from simulations, we will assume that localized photon bubble wave trains grow wherever possible given the local conditions, and saturate at close to the maximum permitted amplitude.  All other factors being equal, the amplitude is limited by the magnetic field, which must be strong enough to resist buckling.  The maximum dimensionless density is $\eta_+ \sim p_m / \bar p_g$, where $p_m$ is the magnetic pressure and $\bar p_g = \rho_0 c_g^2$ is the mean gas pressure (B01; Turner et al. 2005).  In our disk models we will find that other factors may limit the amplitude of the photon bubbles, such as constraints on their wavelengths (which must be shorter than the scale height) or the transition of the disk to a parameter regime in which photon bubble formation is inhibited.  Reassuringly, our models for disk cores will indicate such a high efficiency of radiative transport (where damping is unimportant) that saturation at the maximum permitted levels is unnecessary. 

In the models presented below we will take $\theta \approx 0$, corresponding to a field parallel to the disk equator.  Simulations suggest that $\langle B_\phi^2 \rangle \sim 3 \langle B_\phi B_r \rangle$ (e.g., Hawley 2000), so we will adopt $p_m \sim 3 \alpha p_0$, where $\alpha$ is the usual viscosity parameter and $p_0$ is the radiation pressure.   Simulations in an initially static, nearly horizontal magnetic field (Turner et al. 2005) suggest that the dominant shock trains have the same orientation as the fastest-growing linear mode (Blaes \& Socrates 2003), corresponding to $\phi \approx -70^\circ$ in our notation when radiation damping is unimportant.  We will adopt this value for $\phi$, except where specified; note that the dominant modes are likely to be different when damping and/or finite-$\tau$ effects are important.   
 
\subsection{Photon Bubbles in Accretion Disk Cores}
  
Following B02, we adopt a one-zone model in which most of the disk luminosity is produced within the high-density core of the disk.  We note, however, that the scaling of the damping parameter with height (assuming LTE) is $\xi \propto p^{9/8}/z$, implying that $\xi$ is very large near the equator and that photon bubbles are likely to be inhibited in this region.  The dependence of $\xi$ on the orientation of the photon bubble wave train ($\xi \propto \cos(\theta - \phi) \sim (1-\mu^2)^{1/2}$; eq.~[\ref{morenondim}]) could partially compensate for the increasing effects of damping in the interior of the core.  Here we will assume that most of the energy is generated in the upper half of the core, and adopt $\mu = \sin 70^{\circ}$.  

B02 equations (5) through (12) are unchanged.  B02 equations (2) and (3), which  represent an early attempt to model the undamped cases (a) and (b), must be modified: the term $\bar\rho^2 c_g^4  \kappa^2 / g^2$ in (2) and $\bar p_g ^2   \kappa^2 / g^2$ in (3), which equal $\tau_0^2$, must be replaced by a single power of $\tau_0$.  The correct formula has one power of $\tau_0$, not two, because the wave Mach number $m_p \propto \tau_0^{-1}$ in the low-$\tau$ limit, according to our analysis in \S~2.   B02 incorrectly assumed that the wave speed was unchanged by finite-$\tau$ effects.  In any case, we shall show below that cases (a) and (b) occupy only a small region of the accretion disk parameter space --- therefore, we shall have to replace the entire B02 analysis by a revised treatment based on \S~2 of this paper. 

As in B02, we assume a steady accretion disk and normalize the accretion rate to the Eddington rate, $\dot m \equiv {\dot M / \dot M_E}$, where $\dot M_E = 4\pi GM / \kappa c$ and $M = m M_\odot$ is the mass of the central object.   We normalize $r$ to the gravitational radius, $x \equiv {r c^2  / GM}$, set $\tilde\alpha \equiv \alpha / 0.03$, and define ${\cal D}\equiv 1- (x_{\rm in}/x)^{1/2}$ for the standard assumption of zero torque at the innermost stable orbit, $x_{\rm in}$ (Shakura \& Sunyaev 1973).  We eliminate the dimensionless disk thickness, $\delta \equiv H/r$, in favor of the Eddington factor $\ell$, by using B02 eq.~(7), $\ell = 3\dot m {\cal D}/ \delta x$.  A standard accretion disk, without photon bubbles, has $\ell =1$. The effect of photon bubbles is to increase $\ell$ so that the disk can lose more radiation without thickening. 

As we showed in \S~2, the key parameters determining the properties of photon bubbles are the characteristic optical depth across a gas pressure scale height,  $\tau_0 = \rho_0\kappa r \delta\beta$, and the damping parameter, $\xi = 4 \beta^{1/2}\cos(\theta - \phi)/ M_0 \approx 4\beta^{1/2} \ell \cos(\theta - \phi)/ 3 \alpha$.  Using the formulae from B02 we have
\begin{equation}
\label{tauformula}
\tau_0 \approx 4.3 \times 10^{-5}  \tilde\alpha^{-5/4}(\dot m {\cal D})^{-3} m^{-1/4} \ell^{17/4} x^{33/8} , 
\end{equation}
\begin{equation}
\label{xiformula}
\xi \approx 3.6 \times 10^{-2}  \tilde\alpha^{-9/8}(\dot m {\cal D})^{-1} m^{-1/8} \ell^{17/8} x^{21/16} . 
\end{equation}
Figure 1 shows the parameter space of disk core models with $\ell = \tilde\alpha =1$, $x_{\rm in}=6$, and two values of the central mass.  The upper excluded region corresponds to $\delta > 1$, i.e., disks that are so thick that they must lose mass.  Since this figure assumes $\ell =1$, it corresponds to standard disk models without photon bubbles --- we can use it to determine when photon bubbles may form.  The lowest region corresponds to $\xi > 1$, implying that photon bubbles are inhibited by radiation damping.  The regimes assumed by B02, the undamped cases (a) and (b), occur only in the narrow strips between the curves.  Most of the parameter space that supports photon bubbles lies between the upper curve and the excluded region, and corresponds to case (c) --- critically damped bubbles.    
%\clearpage
\begin{figure*}[t]
\plottwo{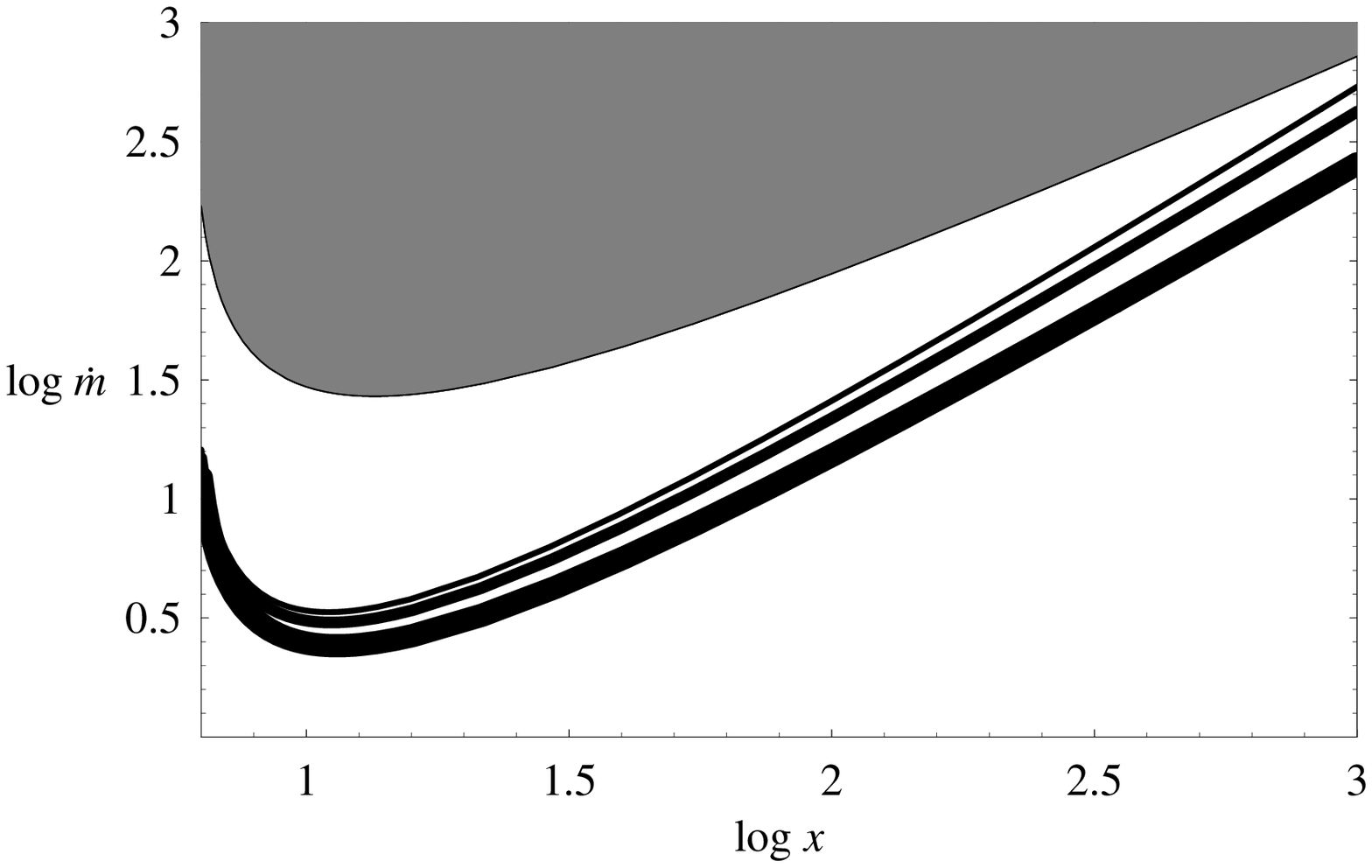}{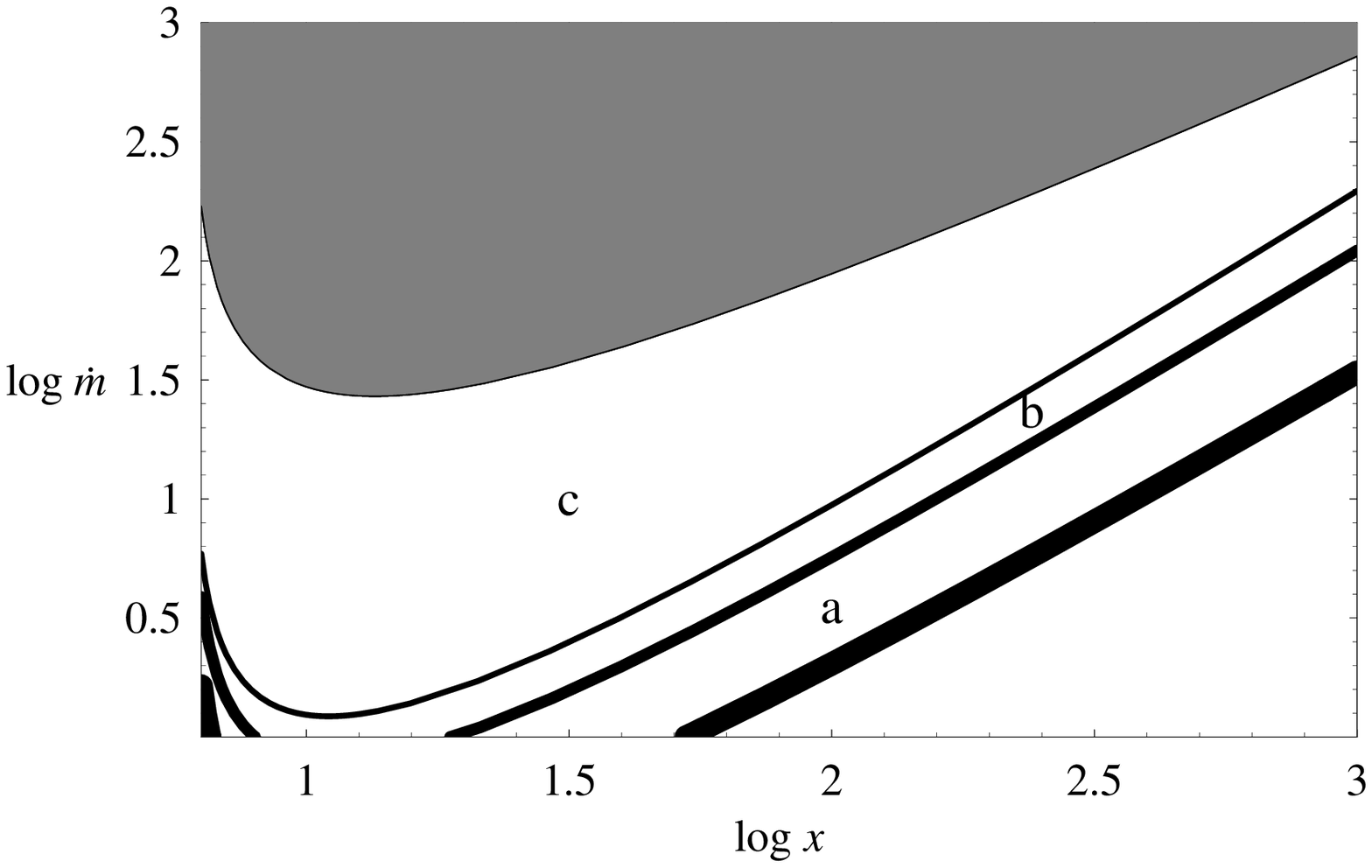}
\caption{Parameter space of thin ($\delta <1$), standard ($\ell = 1$) disk cores with $\tilde\alpha = 1$ and $M = 10 M_\odot$ (left panel); $M= 10^8 M_\odot$ (right panel). The upper shaded region in each plot violates the thinness condition.  The lowest (thickest) curve in each plot corresponds to $\xi =1$, the next curve corresponds to $\tau_0 =1$, and the highest (thinnest) curve corresponds to $\xi = \tau_0$.  Letters on the right-hand panel refer to the photon bubble parameter space discussed in \S~2 and summarized in Table 1. Photon bubbles cannot exist in the region below the lowest curve. Between the upper curve and the excluded region, photon bubbles exist with radiation damping approximately balancing driving forces.  The analysis by B02 is valid only in the narrow strips a and b between curves.}
\label{fig1}
\end{figure*}
%\clearpage
Photon bubbles can exist for parameters such that  $\xi < 1$ and $\delta < 1$.  For fixed values of $m$ and $\alpha$, the size of this region in the $\dot m - x$ plane is a rapidly decreasing function of $\ell$.  We can see this in Fig. 2, which shows how quickly the parameter space available for photon bubbles becomes restricted with increasing $\ell$.  At a given $x$, all boundary curves in the parameter plane move toward higher $\dot m$, but not at the same rate.  The $\delta = 1$ curve and the $\tau_0 = \xi$ curve rise approximately linearly with $\ell$, while the $\xi = 1$ and $\tau_0= 1$ curves increase roughly as $\dot m \propto \ell^2$ and $\propto \ell^{3/2}$, respectively.    
%\clearpage
\begin{figure*}[t]
\plottwo{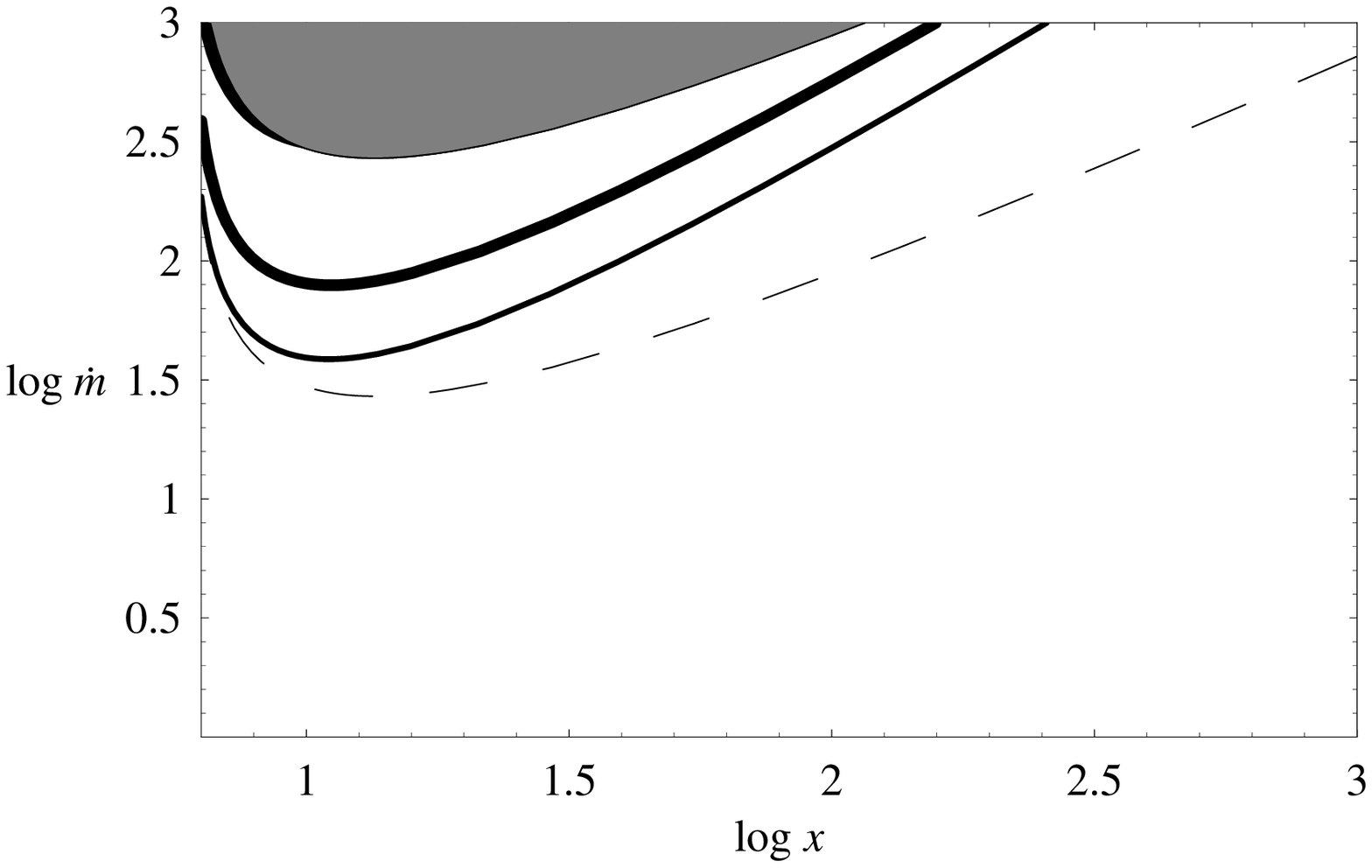}{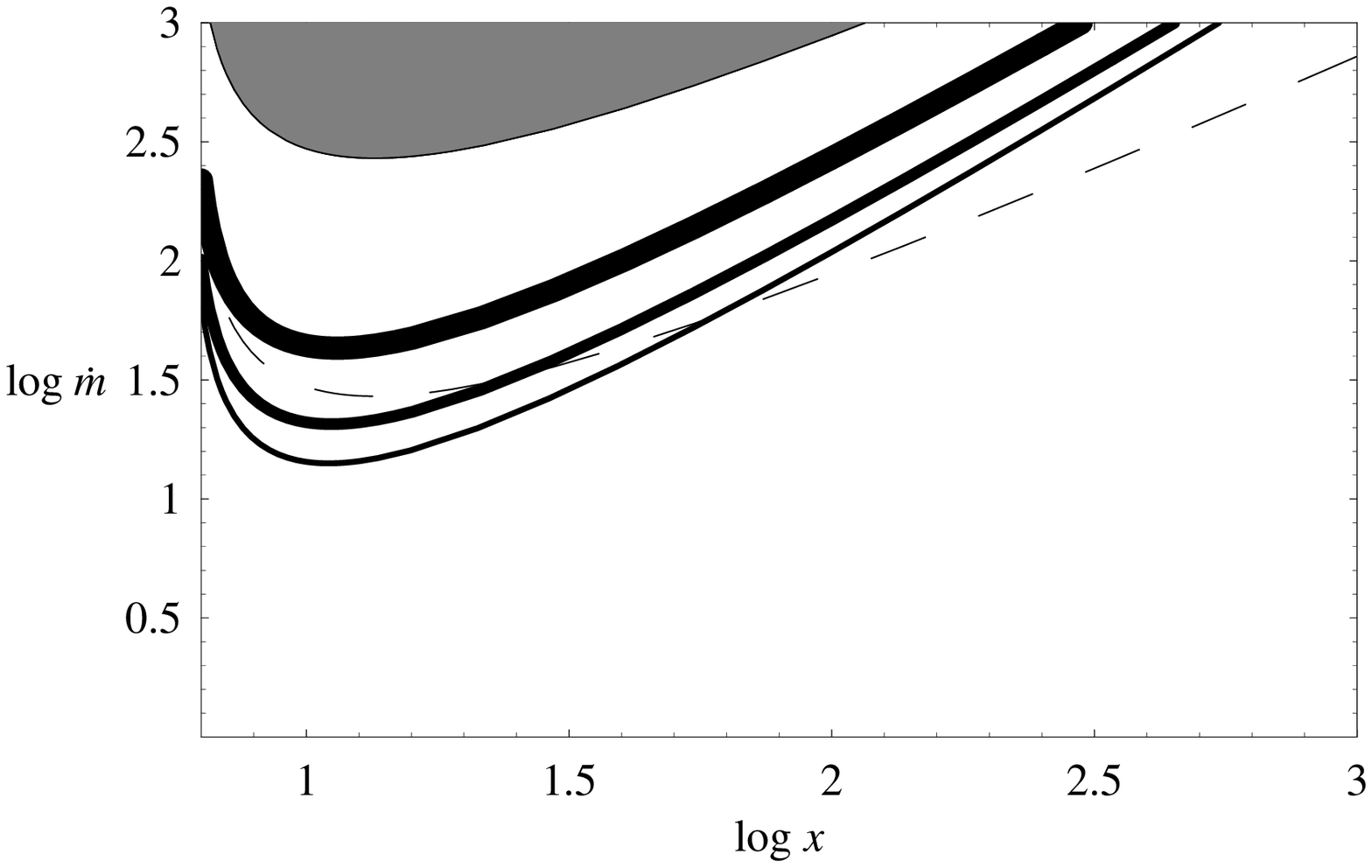}
\caption{Parameter space of thin ($\delta <1$), disk cores with $\tilde\alpha = 1$, $\ell = 10$ and $M = 10 M_\odot$ (left panel); $M= 10^8 M_\odot$ (right panel). The upper shaded region in each plot violates the thinness condition.  The dashed line shows the thinness constraint for $\ell = 1$. The lowest (thinnest) solid curve in each plot corresponds to $\xi =\tau_0 > 1$; the intermediate thickness curve corresponds to $\tau_0=1$.  In the right panel, photon bubbles are possible only between the heavy curve ($\xi = 1$) and the shaded region (case c).  In the left panel the heavy curve has moved into the shaded region, implying that disks with $M=10 M_\odot$, $\ell = 10$ violate the thinness constraint. Note that all boundaries move upward with increasing $\ell$, thus restricting the parameter space available for photon bubbles.}
\label{fig2}
\end{figure*} 
%\clearpage
The maximum possible $\ell$ is therefore reached when the disk conditions approach $\xi \sim 1$, i.e., close to the value of $\ell$ where photon bubble formation is cut off.  This implies
\begin{equation}
\label{lmax}
\ell \la \ell_{\rm max} \approx 5  \tilde\alpha^{9/17}(\dot m {\cal D})^{8/17} m^{1/17} x^{-21/34}.  
\end{equation}
Photon bubbles can affect the disk only where $\ell_{\rm max} >1$, which implies that photon bubbles are important only for 
\begin{equation}
\label{xpb}
x  < x_{\rm pb} \approx 13  \tilde\alpha^{18/21}\dot m ^{16/21} m^{2/21} ,  
\end{equation}
where we have assumed ${\cal D} \approx 1$.  At the same time, the mass flux that can be carried by the disk is constrained by the thinness condition $\delta \leq 1$, which implies 
\begin{equation}
\label{lmin}
\ell \geq \ell_{\rm min} = 1.5 \dot m {\cal D} x^{-1} . 
\end{equation}
Self-consistency demands $\ell_{\rm min} < \ell_{\rm max}$, and therefore that
\begin{equation}
\label{mdmax}
\dot m < 9 \tilde\alpha m^{1/9}{\cal D}^{-1} x^{13/18}  .
\end{equation}

We can place these constraints into a consistent picture for the role of photon bubbles in the central regions of an accretion disk.  
For $\dot m < 3.2 \tilde\alpha^{-9/8} m^{-1/8}$, photon bubbles never form in the disk --- $\xi > 1$ at all radii. For an accretion efficiency of $\epsilon \sim 0.1$, this limit corresponds to disks less luminous than $\sim 0.3 \tilde\alpha^{-9/8} m^{-1/8} L_E$.  Note that photon bubbles set in at lower luminosities (relative to the Eddington limit) for higher mass black holes. If $\dot m > 3.2 \tilde\alpha^{-9/8} m^{-1/8}$, then photon bubbles start to become important at $x \la x_{\rm pb}$, the radius at which the disk first ``hits" the $\xi = 1$ curve.  Since nonlinear photon bubbles tend to evolve toward larger density contrasts, and therefore higher $\ell$, a photon bubble-dominated disk core should have a natural tendency to track along the boundary of marginal stability, corresponding to the curve $\xi =1$. In other words, the tendency for $\ell$ to be as large as possible supplies feedback that drives the disk toward $\xi \sim 1$.  Since the location of the $\xi = 1$ curve in the $\dot m - x$ plane depends on $\ell$, the value of $\ell$ increases with decreasing $x$, lifting the $\xi = 1$ curve and allowing $\dot m$ to remain constant.

The Eddington enhancement factor $\sim 1$ at $x_{\rm pb}$ and increases as $\ell \sim (x/x_{\rm pb})^{-21/34}$ toward smaller radii, provided that the disk is able to sustain the required density contrasts. The geometric disk thickness $\delta \propto (x\ell)^{-1}\propto x^{-13/34}$ at fixed $\dot m$ (and ${\cal D}\approx 1$); therefore the disk thickens with decreasing radius. 

If $\dot m < 170 \tilde\alpha m^{1/9}$ the photon bubble-dominated disk does not violate the thinness criterion at any radius.  Therefore, it may be possible for photon bubbles to liberate radiation at a rate $> 20-130  (\epsilon/0.1) \tilde\alpha L_E$, for black hole masses ranging from $10-10^8 M_\odot$.  Of, course, such supercritical luminosities would drive off mass from the outer layers of the disk where the photon bubble mechanism fails --- we will address the consequences of this mass loss in \S~3.3.  

For higher accretion rates, $\ell$ increases with decreasing $x$ as $\sim (x/x_{\rm pb})^{-21/34}$, until $\delta \sim 1$.  At smaller radii the disk loses mass at a rate sufficient to keep $\delta \sim \xi\sim 1$, i.e., $\dot m \propto x^{13/18}$ where ${\cal D}\approx 1$.  Note that the gravitational energy release still increases with decreasing radius, despite the mass loss. The decline of $\dot m$ continues until $x^{13/18}{\cal D}^{-1}$ is minimized, at $x \approx 17$.  Within that radius the accretion rate levels off at the maximum value calculated above, $\dot m \sim 170 \tilde\alpha m^{1/9}$. 

The foregoing discussion assumes that photon bubbles grow to whatever density contrast is needed in order to keep the disk near the threshold $\xi \sim 1$ and $\ell\sim \ell_{\rm max}$.  Simulations (Turner et al.~2005) suggest that density contrasts in nonlinear photon bubbles tend toward the maximum values permitted by either the magnetic field strength or the wavelength (which cannot exceed the scale height).  For all cases with $\ell \ga 3-6$ (where the threshold depends on $\alpha$ and $m$), we are in photon bubble case (c), with $\tau_0\ll 1$ and $\xi \sim \mu$.  Equation (\ref{Eddfactor}) then gives 
\begin{equation}
\label{Eddfactorc}
\ell \sim 1 + {\mu^2 \over 2 \eta_-}  .
\end{equation}
For reasons mentioned earlier we adopt $\phi = -70^\circ$, so $\mu = 0.94$ and $\ell \sim 0.4/\eta_-$ in the photon bubble-dominated regions of the disk. Since $m_p \approx 1$ and $\eta_+ \eta_- = m_p^2$ from the shock jump condition, the maximum density in the photon bubbles must be $\eta_+ \sim 2.3 \ell $ times the mean density.  

As discussed in the introduction of \S~3, the maximum density is limited by the stiffness of the magnetic field to $\eta_+ \la  3\alpha / \beta$, where $\beta$ is the ratio of mean gas pressure to radiation pressure.  Is this limit on $\eta_+$ consistent with the value required to maintain $\xi = 1$ and $\ell = \ell_{\rm max}$?   For $x_{\rm in} = 6$, half the accretion disk luminosity (for fixed $\dot m$) is produced outside $x_{1/2} = 23.5$, and we will adopt this as the fiducial radius for the purpose of this argument.  At $x_{1/2}$ the maximum accretion rate consistent with the thinness criterion is $\dot m \sim 180 \tilde\alpha m^{1/9}$, corresponding to $\ell \sim 5.6 \tilde\alpha m^{1/9}$ and $\eta_+ \sim 13 \tilde\alpha m^{1/9}$.  Substituting these values of $\dot m$ and $\ell$ into $\beta (x_{1/2}) \sim 0.1 \tilde\alpha^{-1/4} m^{-1/4}\dot m ^{-2} \ell^{9/4}$, we obtain the constraint $\eta_+ < 600 \tilde\alpha m^{2/9}$, which is easily satisfied.  Thus, the strength of the magnetic field is more than adequate to support the required level of density inhomogeneity. 

The other relevant constraint is that the characteristic separation of the dense regions (the photon bubble wavelength) be shorter than the disk scale height, $H$.  From eq.~(\ref{lambdaestimate}), we have $\lambda / H \sim \beta \eta_+ / \sqrt{5} \tau_0 $.  For the maximum accretion rate at $x_{1/2}$ we have $\tau_0 \sim 0.044 m^{-1/9}$, implying $\lambda / H \sim 0.02 \tilde\alpha$.  The dominant scale length of photon bubbles is therefore much smaller than the pressure scale height.   

\subsubsection{Thermal and Viscous Stability}

Domination of radiative transport by photon bubbles does not appear to cure the thermal and viscous instabilities that plague  standard $\alpha$-disks when radiation pressure exceeds gas pressure (Lightman \& Eardley 1974; Shakura \& Sunyaev 1976). Consider a radiation-dominated disk with surface density $\Sigma$ and pressure scale height $H$.  In hydrostatic equilibrium the radiation pressure scales as $p \propto \Sigma H$ while the density obeys $\rho \propto \Sigma/H$ and the gas pressure (in LTE) satisfies $\bar p_g \propto \Sigma^{5/4}/ H^{3/4}$.  Therefore, $\beta \propto \Sigma^{1/4}/H^{7/4}$ and $\xi \propto \Sigma^{9/8} H^{1/8}$. The rate of dissipation per unit area, according to the $\alpha$-model, scales as $Q^+ \propto H^2 \Sigma$, while the rate of energy loss scales as $Q^- \propto \ell H$.

Consider thermal stability first. Since thermal timescales are much shorter than viscous timescales, we can consider $\Sigma$ to be constant.  In the absence of photon bubbles, $\ell =1$ and $Q^-$ increases with $H$ more slowly than $Q^+$, leading to thermal instability (see, e.g., Frank, King \& Raine 2002).  When photon bubbles dominate the energy loss, however, the disk must lie close to marginal stability, $\xi \sim \mu$; otherwise the natural tendency for $\ell$ to be as large as possible would release too much radiation. At constant $\Sigma$, $\xi \sim \tau_H c_g/ c \propto H^{1/8}$ in LTE.  Therefore, a slight increase in $H$ presumably leads to a significant decrease in $\ell$ as $\xi$ increases, thus making the disk less susceptible to photon bubble instability.   Since $Q_+$ increases $\propto H^2$ while $Q_-$ probably decreases (or at least increases more slowly), the disk is thermally unstable.  Note, however, that the dependence of $\xi$ on $H$ is very weak, and depends entirely on the slight increase of gas temperature with radiation pressure under LTE.  The conclusion would be reversed if $\xi$ decreased slightly with increasing $H$, either through a decrease in gas temperature (unlikely), a small decrease in $\Sigma$ (since $\xi$ is much more sensitive to $\Sigma$ than to $H$), or a decrease in the characteristic value of $\mu$.

Now consider viscous instability. According to the $\alpha$-model, the viscous couple satisfies $G \propto pH \propto H^2 \Sigma$.  On viscous timescales, thermal equilibrium is maintained, implying $Q^+ = Q^-$, where $Q^+ \propto H^2 \Sigma $.  When photon bubbles are absent, $Q^- \propto H$ and  thermal equilibrium gives $H \propto \Sigma^{-1}$, implying $G \propto \Sigma^{-1}$.  Since the viscous couple is a decreasing function of surface density, the coefficient of viscous diffusion is effectively negative and the disk is unstable to clumping into rings (Lightman \& Eardley 1974).  When photon bubbles dominate, $\xi \sim$ constant.  The viscous couple decreases rapidly with $\Sigma$, and the disk is unstable. 

Photon bubble transport {\it could} quench both thermal and viscous instability if the density contrasts were close to their maximum values, or scaled with these values.  In this case, $\ell \propto p/ \bar p_g \propto p^{3/4}/\rho \propto H^{7/4} \Sigma^{-1/4}$ in LTE (the LTE assumption is not crucial to the qualitative result).  We therefore have $Q^-\propto H^{11/4} \Sigma^{-1/4}$, cooling increases more rapidly than heating at fixed $\Sigma$, and thermal instability is quenched.  By the same token, in thermal equilibrium the viscous couple increases strongly with $\Sigma$, $G \propto \Sigma^{13/3}$, and the disk is extremely stable on viscous timescales. 

\subsection{Disk Halo}

We characterize the disk halo as the plane-parallel region, sandwiching the core, through which a constant mean radiation flux passes.  We assume that photon bubbles also develop within the halo.  In LTE, we have the scalings $\xi \propto p^{9/8}/z$, $\tau_0 \propto \rho p^{1/4}/ z$, and $\beta \propto \rho /p^{3/4}$ (qualitatively similar to the isothermal scalings $\xi \propto p/z$, $\tau_0 \propto \rho / z$, and $\beta \propto \rho /p$).   

Let us denote core quantities by a subscript ``c" and use a hat to label quantities normalized to core conditions.  Since the flux is constant while the Eddington flux increases linearly with $z$,  the condition of hydrostatic equilibrium requires that $\hat \ell = \ell / \ell_c = (z / H_c)^{-1} = \hat z^{-1}$. 

Although $\xi_c \sim 1$, $\xi$ decreases rapidly with height above the core.  The halo departs from marginal stability and therefore photon bubbles should be able to approach the maximum permitted density contrast.  As the atmosphere becomes increasingly porous, both the density and pressure drop steeply.  This transition zone ends where the maximum allowed value of $\ell$ approaches $\ell_c$. Since the density drops at least as steeply as the pressure, the ordering $\tau_0 \ll \xi$ is preserved and we remain in case (c) of the photon bubble parameter space with $A \sim 1/ \sqrt{5}\tau_0$. The Eddington enhancement factor is still given by eq.~(\ref{Eddfactorc}).  As discussed earlier, the magnetic field strength places an absolute upper bound on the postshock density of $\eta_+ \la 3\alpha / \beta$, and therefore a lower bound on $\eta_-$.  However, this upper bound is superseded by the constraint that the photon bubble wavelength, $\lambda \sim A/\eta_-$ (in units of the gas pressure scale height), not exceed the atmospheric scale length (or a more stringent upper limit associated with coherence length of the magnetic field).  Setting the maximum wavelength to $0.3 \chi z$, where $\chi \le 1$ is a constant parameter intended to mimic the effects of an incoherent magnetic field, we obtain $\ell \sim 0.3 \chi \rho \kappa z$, implying  
\begin{equation}
\label{rhohat}
\hat \rho \sim 10 \chi^{-1} {\alpha \delta_c \over x^{1/2}}   \hat z^{-2}  \ll 1
\end{equation}
in the region above the transition zone.   
 
In scaled variables the equation of vertical hydrostatic equilibrium becomes
\begin{equation}
\label{hydrohat1}
{d\hat p \over d\hat z} = - \hat z \hat\rho  \propto \hat z^{-1}
\end{equation}
at $z \ll r$, implying that $p$ and $\rho$ drop by roughly the same factor in the transition zone, but that $p$ decreases with height logarithmically at higher $\hat z$.  This behavior continues to the ``top" of the halo at 
\begin{equation}
\label{zmax}
\hat z_{\rm max} \sim \min \left[\ell_c, \delta_c^{-1} \right] , 
\end{equation}
where $\delta_c = H_c/r$ is the geometric thickness of the disk core.  If $\ell_c < \delta_c^{-1}$, then the flux passing through the halo at $x$ is sub-Eddington with respect to the mass of the black hole. $z_{\rm max}$ is then the height at which the optical depth across the halo, $\rho (z_{\rm max}) \kappa z_{\rm max}$, declines to $\sim O(1)$, and photon bubbles can no longer be supported.  $z_{\rm max}$ is also the thickness of an ``equivalent" standard accretion disk with the same local flux but no photon bubbles.  The difference between the disk-halo structure dominated by photon bubble transport and a standard accretion disk is that most of the column density in the photon bubble-dominated disk is found in the much thinner core --- only a fraction $\ell_c^{-1}$ of the disk column is found in the upper regions of the halo.  Nevertheless, the scattering photosphere is located near the top of the halo, and therefore the disk will appear to be as thick geometrically as a standard disk of the same luminosity. 

If  $\delta_c^{-1} < \ell_c $, then the halo flux is super-Eddington with respect to the black hole mass, and $z_{\rm max} > r$.  In this case the hydrostatic disk halo is replaced by a radiation pressure-driven wind. We discuss the properties of this wind below.     
 
\subsection{Disk Wind}

If the disk core thickness $H_c$ exceeds $\ell_c^{-1} r$ (i.e., for $\ell_c \delta_c = 3\dot m {\cal D}/2 x$), the disk is super-Eddington and must lose mass.  This condition is equivalent to the halo being geometrically thick, $z_{\rm max} > r$.  In this section we estimate the properties of the wind and show that the mass loss need not be catastrophic.

Treating the outflow is more complicated than treating the static halo because part of the energy flux is advected, $F_{\rm adv} = 4 p v$ (see \S~2.2.4).  Naively, this advective flux seems likely to dominate the energy budget in the wind, leading to such a large mass flux that it would decimate the accretion disk.  To see this, suppose that the halo carries a conserved, vertical wind flux $\dot m_w = \rho v_w$ at $z < r$.  At $z \sim r$, the gravitational acceleration switches from increasing with height $\propto z$ to decreasing $\propto r^{-2}$.  Let us assume that the flow reaches escape speed near this point, $v_w \sim v_K =(GM/r)^{1/2}$, where $v_K$ is the Keplerian speed at $r$. We saw in \S~3.2 that the static halo (above the transition layer) has roughly uniform radiation pressure, $p\sim 10 \chi^{-1} p_c \alpha \delta_c x^{-1/2}$ (to within a logarithmic factor).  Substituting this pressure into the expression for $F_{\rm adv}$ and setting $v = v_K$, we find that $F_{\rm adv} \sim 13 \chi^{-1}x^{-1/2}(g_c c/ \kappa) \ell_c $, where $g_c$ is the gravitational acceleration at the top of the disk core.   The total radiation flux passing through the halo, $F_{\rm tot}$, is equal to $(g_c c /\kappa) \ell_c $.  Under these assumptions we would therefore conclude that most of the energy --- at least in the inner parts of the disk, say, near $x_{1/2} \sim 23$ where most of the energy is generated --- is advected.  If the advective flux dominates then the radiation becomes trapped in the outflow and the mass loss is maximized.  The mass loss rate per unit disk area is then $\dot m_w \sim F_{\rm tot} / v_K^2$, which is just enough to prevent the total luminosity from exceeding the Eddington limit by a substantial factor.  

However, this argument appears not to be self-consistent.  According to our prescription for photon bubble transport, the diffusive component of the flux is given by $F_{\rm diff} = - (c/\kappa) (\nabla p)/\rho) \ell \sim - 0.3\chi c z \nabla p $.  Unless $\chi$ is small, at $x\gg 1$ the diffusive flux should dominate over the advective flux if $z |\nabla p| \sim O(p)$.  The reason we found that the advective flux could dominate in the halo solution is that, according to this solution, $z |\nabla p| \ll O(p)$.  But as soon as the advective flux becomes appreciable, the acceleration increases and one can show from the equation of motion that this ordering no longer applies.  We conclude that the flow cannot become strongly advection-dominated.  

Ignoring advection allows us to construct a simple, quasi-one-dimensional model for the flow dynamics.  Defining $y \equiv z/r$, we mimic the transition from vertical to radial flow by defining the conserved mass flux per unit disk area as $\dot m_w = \rho(y) v(y) (1 + y^2)$.  By the same token, we write the radiation flux as $F_{\rm tot} \sim F_{\rm diff} \sim (g_c c/\kappa) \ell_c (1 + y^2)^{-1}$ and the effective gravity as $(v_K^2/r) y/(1+y^2)^{3/2}$.  If we further define $w \equiv v/ v_K$, $Q \equiv \delta_c \ell_c v_K/ (0.3 \dot m_w \kappa r)$, and denote differentiation with respect to $y$ by a prime, we obtain the equation of motion
\begin{equation}
\label{mdotmotion}   
w w' = Q {w\over y} - {y\over (1+ y^2)^{3/2}} ,
\end{equation}
which is valid where $\ell > 1$.  This equation reproduces the structure of the static halo in the subsonic zone at $y \ll 1$, with $\rho \propto z^{-2}$ implying $v\propto z^2$. 

Although this equation is tricky to solve exactly, because of the singularity, it is straightforward to show that sensible solutions exist only for $Q \sim O(1)$. For $Q \gg 1$ (small $\dot m_w$), the inertial term on the left-hand side of eq.~(\ref{mdotmotion}) is negligible for all $y < 1$.  The flow never reaches escape speed and decelerates at large $y$.  For $Q \ll 1$ (large $\dot m_w$), the inertial terms become appreciable at $y \sim Q \ll 1$, but because the porosity due to photon bubbles (as measured by $\ell$) decreases with decreasing density, the flow never receives enough momentum to accelerate it to escape speed --- again the flow decelerates.  Only for $Q \sim O(1)$ does the flow receive the right momentum to accelerate it to $\sim v_K$ at $y\sim 1$, at which point it can coast (with possibly some additional mild acceleration) to large $r$. 

The mass loss rate per unit disk area is then given by 
\begin{equation}
\label{mdotw}
\dot m_w \sim {3 \over 8\pi } Q {\dot M_E \over r_g^2} {\dot m {\cal D} \over x^{5/2}} .
\end{equation}
Mass loss begins at an outer radius $x_0$ where the accretion rate $\dot m_0$ first exceeds the Eddington limit, $\dot m_0 \sim 2 x_0 /3$. For the coefficients obtained above (taking into account mass loss from both sides of the disk and assuming that $\dot m$ does not exceed the limit imposed by photon bubble transport), mass flux is depleted by the wind according to the equation $d\dot m / dx \approx 1.5 Q ({\cal D} / x^{3/2}) \dot m$.  For $x_{\rm in} = 6$ the solution is 
\begin{equation}
\label{movermzero}
{\dot m (x) \over \dot m_0} \ga \exp\left\{ Q \left[ 3.7 \left(x^{-1} - x_0^{-1}\right) - 3 \left(x^{-1/2} - x_0^{-1/2}\right)\right] \right\} .
\end{equation}
For $Q = 1$ and $x_0 \la 1000$, more than 50\% of the accreting matter reaches the center, implying that highly super-Eddington luminosities are possible without destroying the accretion disk.  This result should be regarded strictly as illustrative, because the coefficient in eq.~(\ref{mdotw}) is very approximate and could be off by a factor of several in either direction.  Moreover, the fact the $Q \sim 1$ suggests that $F_{\rm adv}$ is at least comparable to the diffusive flux.  If $F_{\rm adv}$ is a sizable fraction of $F_{\rm tot}$ then the diffusive flux is reduced by a factor $1 - F_{\rm adv}/F_{\rm tot}$.  The same factor would multiply the first term on the right-hand side of eq.~(\ref{mdotmotion}).  Acceptable values of $Q$ would be increased by the reciprocal of this factor, and the mass loss rate would be correspondingly diminished.  This argument should be valid provided that the total radiative luminosity is approximately conserved through the outflow, i.e., if only a small fraction of the radiation flux is converted to kinetic energy.

Even if the mass loss is not catastrophic, the fact that it is a substantial fraction of the accretion rate means that the total kinetic luminosity of the wind. $L_w$, is likely to be only a few times smaller than the radiative luminosity of the disk.   For super-Eddington luminosities this can represent a large flux of energy, which could dramatically affect the environment.

\subsection{Disk Temperature}

 Normal accretion disks have color temperatures that exceed their effective temperatures because the radiation pressure at the ``true" photosphere, where the escaping radiation is thermalized, is larger than the radiation pressure at the scattering photosphere.  Typical color corrections, $f_{\rm col} \equiv T_{\rm color}/ T_{\rm eff}$, lie in the range $\sim 1.5-2$ (Shimura \& Takahara 1995; Davis et al. 2005), implying that thermalization occurs at a scattering optical depth $\sim 5-20$.  How does photon bubble transport affect disk color? 

A detailed treatment of this issue is beyond the scope of this paper, but we are able to make some comments about the expected trends.  First, we note that the absorption opacity used to estimate thermalization effects should be evaluated at a density close to the maximum density found in the photon bubbles, $\eta_+ \rho_0$, where $\rho_0$ is the mean density. Although the densest gas occupies only a fraction $\sim \eta_+^{-1}$ of the volume, the absorption opacity $\kappa_{\rm abs}$ is proportional to $\eta_+$.  Therefore the effective opacity is $\kappa_{\rm eff} \sim [\eta_+ \kappa_{\rm es}\kappa_{\rm abs}(\rho_0)]^{1/2}$ where $\kappa_{\rm es}$ is the electron scattering opacity.  The absorption opacity (presumably a mixture of free-free and bound-free opacities) generally decreases with radiation pressure in LTE --- for free-free absorption we have $\kappa_{\rm ff} \propto p^{-7/8}$, with a somewhat slower mean decline if bound-free transitions are important. 

First consider the disk core.  Compared to a standard disk at radius $x$ with the same $\dot m$ and $\alpha$, the photon bubble-dominated disk has both higher pressure and higher density, $p_c\propto \ell_c$ and $\rho_c \propto \ell_c^3$. We also have $\eta_+ \propto \ell_c$, therefore the effective opacity is higher than in the equivalent standard disk by a factor $\ell_c^{25/16}$.  The effective optical depth for thermalization scales as $\tau_{\rm eff} \propto \rho \kappa_{\rm eff} H_c \propto \ell_c^2 \kappa_{\rm eff} \propto \ell_c^{57/16}$.  Thus, thermalization should be much more effective in photon bubble-dominated disk cores than in standard accretion disks, mainly because the typical densities are much higher.   

Because the density and pressure drop by roughly the same factor between the core and halo, effective opacities ($\propto \rho^{1/2} p^{-7/16}$) are unlikely to change by a large factor across the transition region.  However, the large density drop (by a factor $\sim 10 \alpha \delta_c / x^{1/2}$) implies a substantial decrease in the effective optical depth. Whether the radiation is still thermalized in the lower parts of the halo therefore depends on details.  Toward the upper halo, the pressure is roughly constant, $\rho \propto z^{-2}$, and according to our estimates for case (c) photon bubble transport, $\eta_+ \propto z$.  Therefore, the effective optical depth decreases $\propto z^{-3/2}$.  Thermalization will certainly fail well below the top of the halo, where the scattering optical depth approaches unity.  These general arguments should also apply to the wind, in the super-Eddington case. If thermalization is still effective in the lower parts of the halo, the color correction should be close to unity, i.e., photon bubble-dominated disks would be cooler than their standard counterparts of the same luminosity, by a factor of $\sim 1.5-2$.  

The higher luminosities of super-Eddington disks, of course, would offset a smaller color correction.  Roughly, we expect $T_{\rm color} \propto f_{\rm col }M^{-1/4} (L/L_E)^{1/4}$.  Therefore, a $10M_\odot$ black hole radiating at $10 L_E$ with no color correction would still appear hotter than a $100 M_\odot$ black hole radiating at the Eddington limit with $f_{\rm col} = 2$, but only by a factor of 1.6.  If some thermalization occurred further out in the wind, or if radiation trapping were important in the outflow, then the color temperature of the low mass hole could be smaller.

\section{Summary and Discussion}

\subsection{Principal Results}

We have attempted to refine and flesh out our earlier treatment (B02) of the possible role of photon bubbles in accretion disks.  The speculative nature of this work cannot be overemphasized, especially given the fact that it is unknown whether highly nonlinear photon bubbles develop in the presence of hydromagnetic turbulence generated by MRI.  We are motivated to take an optimistic view by the fact that photon bubbles develop in uniform fields on timescales much shorter than the timescales that characterize MRI (Gammie 1998; B02; Turner et al. 2005).  Moreover, the fact that MRI appears to predict a predominantly horizontal magnetic field suggests that field coherence may be adequate to support photon bubbles of moderately long wavelength (and therefore of large density contrast). Nevertheless, the time-dependence of the field, as well as its small-scale structure, could very well prove fatal to the premises upon which this paper is based. At least, it is hoped that the arguments presented in this paper will help to frame future radiation hydrodynamical simulations, which will settle these questions. 

\subsubsection{Photon Bubble Theory}

Our analysis is based on an analytic model of photon bubbles as a periodic train of planar, isothermal shocks mediated by gas pressure (B01).  The magnetic field is sufficiently stiff that it enforces one-dimensional motion.  The general validity of this model has been verified by numerical simulations (Turner et al. 2005).  However, photon bubbles in accretion disk cores occupy a part of parameter space that had not been analyzed previously.  

First, the optical depth across low density regions of the photon bubble, as measured by the parameter $\tau_0$ --- the optical depth for gas at the mean density, across the {\it gas} pressure scale height --- is predicted to be small.  The usual diffusion law relating flux to the radiation pressure gradient and the density has to be modified, since the flux cannot adjust to changes in density across distances shorter than a photon mean free path.   We therefore generalized the closure to the radiation hydrodynamics equations described by Ruszkowski \& Begelman (2003), to take account of fluid motions (see Appendix A).  This generalization greatly complicated the equation describing the photon bubble structure (\S~2), changing it from a first- to a third-order nonlinear differential equation.  Nevertheless, the generic flow pattern, with a critical point at some characteristic flow speed, is preserved, and we were able to approximate the principal features of photon bubbles at finite and low optical depths.  Most importantly, we found a tendency for the low-density parts of the photon bubble to ``fill in," so that the optical depth across any scale length never becomes much smaller than one.  Another new behavior, which appears as long a radiation damping is sufficiently weak, is an increase in the photon bubble phase speed to values well above the gas sound speed. 

Second, radiation damping is predicted to be important in accretion disks.  A damping parameter $\xi$ --- essentially the ratio of the radiation diffusion time to the sound crossing time in the gas --- was included in the B01 models, which apply (with some minor corrections: see Turner et al. 2005) for $\tau_0 \gg 1$.  We extended the analysis to include the effects of damping at finite and small optical depth.  Although we were not able to provide a rigorous proof, it appears that photon bubbles with large density contrasts can form only when $\xi \la 1$ or, more precisely, when $\xi$ is less than some value which depends on the orientation of the shock fronts and the magnetic field but which is probably $\sim O(1)$ in practice.  As in the case of photon bubbles at large $\tau$, the ``Eddington enhancement factor" $\ell$ --- which is the ratio of the flux to the {\it local} Eddington flux, $gc/\kappa$, and not to the global Eddington limit --- is given essentially by the ratio of the mean density to the minimum density, as first pointed out by Shaviv (1998). 

\subsubsection{Accretion Disk Theory}
 
We applied our simple photon bubble theory to a one-zone $\alpha-$model for the equatorial region of the disk around a black hole. The treatment resembles standard models for radiation pressure-supported disks except for the inclusion of $\ell$ as a free parameter.  Where photon bubbles do not exist, $\ell = 1$.  In the absence of photon bubbles, $\xi$ increases with $r$, implying that damping prevents the formation of photon bubbles beyond a certain radius.   For accretion disks less luminous than about $0.3 (\alpha / 0.03)^{-9/8} (M/M_\odot)^{-1/8} L_E$ (for an accretion efficiency of $0.1$), $\xi > 1$ at all radii and photon bubbles cannot form. 

For more luminous disks, the onset of radiative transport by photon bubbles, at the radius where $\xi$ first drops below $\sim 1$, causes $\ell$ to increase with decreasing $r$.  In a steady state with constant $\dot m$, unchecked energy loss via photon bubbles would increase $\xi$ to the point where photon bubble formation shuts off again.  We therefore deduce that there is a feedback mechanism, which should regulate the amount of energy transported by photon bubbles so that $\xi \sim 1$. The photon bubble density contrasts required to maintain such an equilibrium are small compared to the maximum values permitted by photon bubble theory, lending credence to the conjecture that such contrasts could be attained under turbulent conditions. 

This feedback mechanism, which should operate on a dynamical timescale, makes specific predictions about the thermal and viscous stability of disks dominated by photon bubble transport.  Under the assumptions of an $\alpha-$model viscosity, we find that the steady state solution with $\xi \sim 1$ does not cure the well-known thermal and viscous instabilities that plague radiation pressure-dominated disks.  An alternative outcome, in view of this instability, is that there is no steady accretion.  Instead, the inner disk might cycle through a sequence of states in which the surface density builds up at small $\xi$, where the disk is thin and viscous transport inefficient, and is subsequently depleted by accretion or evaporation of a thickened disk, without photon bubbles, at large $\xi$.  Recall from \S~3.1.1 that both thermal and viscous instabilities {\it can} be quenched by photon bubble transport if $\xi \ll 1$ and the photon bubbles are able to approach the maximum permitted density contrast. 

Steady state disk cores with $\xi \sim 1$ can liberate super-Eddington luminosities via photon bubble transport.  This result is qualitatively and quantitatively similar to the conclusion we reached in B02, although the details of the photon bubble transport model, especially the feedback effect, are different.  The predicted maximum luminosity depends weakly on mass, with $L_{\rm max} \sim 20 (\alpha/0.03) (M/M_\odot)^{1/9} L_E$.

Super-Eddington luminosities necessarily drive winds.  As the disk density declines, its ability to support photon bubbles with large density contrasts decreases.  As $\ell$ decreases, the mean force exerted by the escaping radiation increases, driving the wind.  In particular, if the advected flux $4pv$ exceeds the diffusive flux, then the mass loss will probably decimate the accretion disk, decreasing $\dot m$ at such a rate ($\dot m \propto r$) that the luminosity cannot significantly exceed $L_E$.  Our approximate analysis suggests that the diffusive flux probably prevails, allowing most of the accreted matter to reach the inner disk.  Although this argument is tentative and will require detailed confirmation, it suggests that super-Eddington luminosities need not be compromised by mass loss.  Mass loss could also be reduced if gas becomes trapped on closed flux loops anchored in the disk (B02; Socrates \& Davis 2005).  

\subsection{Caveats}

We have already emphasized the overwhelming caveat of this paper: that the incoherence and time-dependence of the magnetic field might prevent the formation of photon bubbles in accretion disks altogether.  In this section we describe a number of other significant uncertainties related to the arguments presented.

1.  We used a highly approximate radiative transfer model to estimate the effects of finite optical depth on photon bubbles.  The closure proposed in RB was tested against Monte Carlo calculations of radiative transfer through plane-parallel slabs.  This test provides a reasonable facsimile of conditions in photon bubbles, and gives reasonable agreement (although the agreement is somewhat worse after the correction of an error in RB; see Appendix A)  However, for the present problem we were required to generalize this closure to include the effects of motion.  No comparable tests of been done of the generalized model.

2. We did not solve the full, third-order nonlinear differential equation for photon bubble structure in the presence of finite optical depths.  The full equation is two-point boundary value problem with a critical point and eigenvalue, in which the boundary conditions depend on each other and on the eigenvalue.  To obtain a quick estimate of the functional dependence of the solutions on parameters, we devised an approximate method to find asymptotic solutions close to solutions that cross the critical point. We have no way of knowing how accurate this scheme is, although it gives reasonable (though not perfect) results in the high-$\tau_0$ limit.   We were also unable to prove that no high-contrast solutions exist for large $\xi$, although the approximate method suggests that this is the case.   

3. We adopted a simple, one-zone model for the disk core structure, assuming that the density scale height coincides with the thickness of the dissipation layer at each radius.  However, simulations suggest that much of the dissipation may occur outside the density core (Miller \& Stone 2000; Turner 2004; Hirose et al.~2005).  If the bulk of the dissipation occurs at high optical depth, but outside the density core, then $\xi$ will be systematically lower than our estimates indicate.  Photon bubble transport could then be important at larger radii and larger accretion rates.  In this sense, our one-zone model is a ``worst-case scenario" for assessing the potential importance of photon bubble transport.  On the other hand, if most of the dissipation occurs at such low optical depths that photon bubbles cannot form (e.g., Socrates \& Davis 2005), then the issue of photon bubble transport is moot.  

4. A related issue concerns the likely absence of photon bubbles deep within the disk core.  $\xi$ is a monotonically decreasing function of height, formally blowing up at the equator where the vertical gravity vanishes.  This suggests that in a disk core with $\xi_c \sim 1$, photon bubbles could exist only in the surface layers.  What happens if a sizable fraction of the dissipation occurs near the equator, where photon bubble transport is absent?  If this energy were carried to the photon bubble-dominated zone by radiative diffusion, it would puff up the disk core to the point that super-Eddington luminosities would be impossible.  However, this is probably not what happens.  The existence of the photon bubble-dominated zone acts like a drop in opacity near the surface, which causes the specific entropy to decrease with height in the disk interior and can drive strong convection down to the equatorial zone. We demonstrate this effect explicitly in Appendix B, by constructing and solving a toy model. This model shows that the ratio of convective to diffusive flux is determined by effective porosity of the disk near the surface; for photon bubble transport the ratio is $\sim \ell \gg 1$. As long as the viscosity parameter $\alpha$ is $\ll 1$, the required convective motions are highly subsonic and the convection can be efficient (see, e.g., Blaes \& Socrates 2003; Turner et al.~2005).  There is no problem with convection transporting a super-Eddington flux.  The damping parameter  $\xi$ is also affected by the orientation of the shock fronts, represented by the parameter $\mu$ with $0< \mu <1$. We arbitrarily set $\mu $ to a fixed value associated with the fastest growing linear photon bubble modes in the high-$\tau_0$ limit.   Although this assumption seems consistent with large-$\tau_0$ simulations at low $\xi$ (Turner et al. 2005), it is possible that the dominant orientation would be affected by damping, or that $\mu $ would adjust to local conditions so that $\xi$ remains small over a much wider range of conditions than we have assumed. 

5. Likewise, we adopted a very simple prescription for the effects of photon bubbles in the halo and wind, assuming that $\ell$ would be close to the maximum permitted value where $\xi < 1$, but would rapidly decrease with $\xi $ for $\xi \ga 1$.  This was the basis for the proposed feedback mechanism, but it ignores the dependence of damping effects on the relative orientation of the magnetic field and the photon bubble shock fronts.   Also, we adopted a simple prescription for determining the maximum $\ell$, based on the maximum allowed photon bubble wavelength or the maximum density contrast permitted by the magnetic field strength.    

6. Finally, we modeled only the density inhomogeneities due to photon bubble shock trains propagating in a static magnetic field, under the influence of the vertical component of gravity. There are other ways of creating inhomogeneities in a radiation-dominated accretion disk, and since enhanced radiation transport is a generic characteristic of highly inhomogeneous atmospheres (Shaviv 1998), we may have underestimated the disk porosity.  In particular, the accelerations associated with MRI would create transient zones of artificial gravity that could lead to the formation of photon bubbles close to the disk equator.  Accelerations associated with magnetic tension on curved field lines --- also driven by MRI --- would create density inhomogeneities via a mechanism akin to the Parker (1966) instability.  Inhomogeneities with density contrasts of order the ratio of the magnetic pressure to the mean gas pressure are seen in radiation--MHD shearing-box simulations (Turner, Stone, \& Sano 2002; Turner et al.~2003; Turner 2004); the extent to which these result from photon bubbles remains to be determined.

\subsection{Astrophysical Consequences}
 
In B02, we discussed several applications of photon bubble-dominated accretion disks, most notably, ultraluminous X-ray sources (ULXs: Fabbiano 1989; Colbert \& Mushotzky 1999; Zezas \& Fabbiano 2002) and rapidly accreting supermassive black holes.  Here we briefly amplify on these suggestions, based on our new results.

It remains controversial whether ULXs are intermediate-mass black holes radiating at sub-Eddington luminosities, or stellar-mass black holes radiating at a few times $L_E$ (or whether there are two classes: Watarai et al. 2005). Our new analysis suggests that several aspects of photon bubble-dominated disks may make it difficult to distinguish these cases.  At a given luminosity, the higher color temperatures associated with smaller disks around lower mass black holes would tend to make super-Eddington disks appear hotter. Some ULXs appear hotter than typical black-hole accretion disks (Makishima et al. 2000; Mizuno et al. 2001) while others appear cooler (Miller, Fabian, \& Miller 2004).  However, we saw in \S~3.4 that the color corrections of photon bubble-dominated disks could be smaller than those of standard disks.  This would partially compensate for the smaller surface area, although the super-Eddington disks probably would still be warmer.  Radiation trapping effects --- although we expect them to be modest --- would work in the same direction, with photon energies adiabatically shifted to lower values.  Another possibility is that the accretion rate at large $r$ exceeds the maximum allowed by photon bubble transport.  Such disks could also produce super-Eddington luminosities, but with $\dot m$ decreasing toward smaller $r$ according to eq.~(\ref{mdmax}).  The energy output would be much less centrally concentrated than in a mass-conserving disk, and the spectrum should be considerably cooler.  In this case, the mass loss would be driven directly from the thick disk, without the intermediary halo.  Radiation trapping could be much more important, which could reduce the color temperature further.  We note that a similar model was proposed by Watarai, Mizuno \& Mineshige (2001).  Finally,we point out that the winds from super-Eddington disks would have scattering photospheres at radii several times larger than their points of origin.  This could smear out variability at the high frequencies associated with QPOs from stellar-mass black holes, although lower frequencies would not be affected (Strohmayer \& Mushotzky 2003).

Super-Eddington accretion could lead to the very rapid growth of supermassive black holes at high redshifts.  Accretion rates could exceed the Eddington value by 1--2 orders of magnitude (depending on black hole mass), leading to e-folding times shorter than the Salpeter time by 1--2 orders of magnitude, i.e., $<4$ million years for seed black holes of stellar mass, decreasing to a few hundred thousand years for $M > 10^6 M_\odot$.  Although we have argued that mass loss need not decimate a super-Eddington accretion disk, the wind would probably carry an appreciable fraction of the luminosity in kinetic energy, and would therefore have a substantial effect on the surrounding protogalaxy.  The release of so much energy on such a short timescale (shorter than the dynamical time of all but the innermost regions of the galaxy) would be like setting off a bomb in the middle of the protogalaxy, rather than a quasi-steady wind.

\acknowledgments
I am indebted to Ari Socrates and an anonymous referee for comments that helped to improve the paper. This research was supported in part by NASA Astrophysical Theory Program grant NAG5-12035 and by the National Science Foundation under Grants AST-0307502 and PHY-990794.  Part of this work was carried out at the Kavli Institute for Theoretical Physics at the University of California, Santa Barbara; I thank the members of KITP for their hospitality. 

\appendix

\section{Radiative Transfer in a Moving, Inhomogeneous Atmosphere}

In this appendix we generalize the radiative transfer model developed by Ruszkowski \& Begelman (2003; RB) to handle motion in an optically thick atmosphere containing large, localized density gradients.  We also correct an error in RB. The approximation is intended to apply where the overall optical depth through the atmosphere is large, but where the optical depth across a density gradient length scale may be small.

We approximate the intensity in the frame comoving with the gas as  
\begin{equation}
\label{intensity}
I(\mathbf{x},\hat{\Omega})=I_{0}(\mathbf{x},\hat{\Omega})+
\frac{3}{4\pi}\hat{\Omega}\cdot\mathbf{F}(\mathbf{x}),
\end{equation}
where $\mathbf{F}$ is the local flux vector in the comoving frame, $\hat{\Omega}$ is the
directional unit vector, and $\mathbf{x}$ is the position vector. 
We assume that all the odd moments of $I_{0}$ vanish, i.e., 
$\int{\Omega}_{i}I_{0}d\Omega=\int{\Omega}_{i}{\Omega}{}{}_{j} {\Omega}_{k}I_{0}d\Omega=0$.  To $O(v/c)$, the equation of radiative transfer for a scattering atmosphere may be written 
\begin{equation}
\label{radtrans}
\frac{1}{\sigma c} {DI \over Dt} + \frac{1}{\sigma}\hat\Omega \cdot\nabla I + 4 {I\over \sigma c} \hat \Omega \cdot \nabla (\hat\Omega \cdot {\bf v}) =-I+\frac{1}{4\pi}\int I d\Omega,
\end{equation}
\noindent
where $\sigma=\rho\kappa$ is the scattering coefficient and $DI/Dt$ is the Lagrangian time-derivative of the intensity (Mihalas \& Mihalas 1984). The third term on the left-hand side comes from the Doppler shift and contains a factor $4$ instead of  $3$ because $I$ is the integrated over frequency. Taking the zeroth moment of eq.~(\ref{radtrans}) we recover the usual energy equation, eq.~(\ref{radentropy}).  The components of the first moment are 
\begin{equation}
\label{divTmoment}
F_i=- {1\over \sigma c}{D F_i \over Dt} -\frac{c}{\sigma}(\nabla\cdot\mathbf{T})_i - {4\over \sigma c }{3\over 4\pi} \int 
\Omega_i (\hat\Omega\cdot {\bf F}) \hat\Omega\cdot\nabla(\hat\Omega\cdot {\bf v}) d\Omega 
\end{equation}
where $\mathbf{T}$ is the radiation stress tensor, the components 
of which are 
\begin{equation}
\label{Tijdef}
T_{ij}=\frac{1}{c}\int{\Omega}_{i}{\Omega}_{j}I_{0}d\Omega .
\end{equation}
The closure relation for $T_{ij}$ in terms of $F_{i}$ and $J\equiv \frac{1}{4\pi}\int Id\Omega$ 
can be obtained by calculating the second moment 
of eq.~(\ref{radtrans}), assuming the form of the intensity given by eq.~(\ref{intensity}). This leads to
the equation for the radiation stress tensor:
\begin{equation}
\label{Tij}
T_{ij} =  \frac{u}{3} \delta_{ij} -\frac{1}{5\sigma c}\left(\frac{\partial F_{i}}{\partial x_{j}}+\frac{\partial F_{j}}{\partial x_{i}}\right) - {1\over \sigma c}{DT_{ij} \over Dt} - {4\over \sigma c} \int \Omega_i \Omega_j I_0\  \hat\Omega\cdot\nabla(\hat\Omega\cdot {\bf v}) d\Omega , 
\end{equation}
where $u = 4\pi J/c$ is the energy density and the radiation pressure is $u/3$.  The Eddington factor is 1/3 because the overall optical depth of the atmosphere is large.  Using eq.~(\ref{Tij}) in eq.~(\ref{divTmoment}) (with the assumption that $I_0$ is isotropic) then gives a set of second-order partial differential equations for the components of $\mathbf{F}$, which we use in \S~2.

Note that the second term on the right-hand side of eq.~(\ref{Tij}) differs from RB eq.~(6), which we believe to be in error.  This error propagated into RB eq.~(7), and thence into the slab models used for comparison with Monte Carlo simulations.  The function $\alpha^2 (\mu)$ in these models should simply be the factor 5.  We note that analytic fits using the corrected formulae are still reasonable, although fits using 10 instead of 5 appear to be much better.  We use 5 in the expressions presented in this paper. 

The last term on the right-hand side of eq.~(\ref{Tij}) represents radiation viscosity. If the optical depth across any flow structure, $\Delta\tau$, satisfies $\Delta\tau \ll c/v$, then we can neglect this term as well as $D{\bf F}/Dt$ and the final term in eq.~(\ref{divTmoment}), the $D{\bf T}/Dt$ term in eq.~(\ref{Tij}), and the ``radiation trapping" ($Dp/Dt$) term in eq.~(\ref{radentropy}).  This approximation is relevant in accretion disks, hence we use  
\begin{equation}
\label{radentropy2}
\nabla \cdot {\bf F} + 4 p \nabla\cdot {\bf v} =0  
\end{equation}
\begin{equation}
\label{divTmoment2}
{\bf F} = -\frac{c}{\sigma}(\nabla\cdot\mathbf{T})  
\end{equation}
for the energy and radiative transfer equation, respectively, where 
\begin{equation}
\label{Tij2}
T_{ij} =  p \delta_{ij} -\frac{1}{5\sigma c}\left(\frac{\partial F_{i}}{\partial x_{j}}+\frac{\partial F_{j}}{\partial x_{i}}\right).   
\end{equation}

\section{Convective Transport in a Photon Bubble-Dominated Disk: A Toy Model}

Generic models of radiation pressure-supported disks tend to be convectively unstable.  This tendency is exacerbated if there is a layer at high $z$ where photon bubbles form, since this has the same effect as introducing an opacity that decreases strongly with height. Since the radiation pressure gradient is determined by hydrostatic equilibrium, and the divergence of the radiation flux is fixed by the dissipation rate, a sudden decrease in the opacity leads to a strong entropy inversion. 

To estimate the magnitude of this effect, consider the vertical structure of a radiation pressure-dominated slab in the presence of gravity $g(z) = -z$ and radiative diffusivity $c/\kappa = \chi(z)$.  Since we will only be interested in ratios of fluxes, we will use convenient nondimensional parameters.  The presence of photon bubbles at large $z$ but not near the equator corresponds to $d\chi/dz >0 $. Assume that the emissivity is proportional to density and that the total flux can be decomposed into a diffusive part and a convective part, $F = F_d + F_c$.  The vertical structure equations are:
\begin{equation}
\label{B1}
p' = -\rho z
\end{equation}
\begin{equation}
\label{B2}
F' = F_d' + F_c' = \rho
\end{equation}
\begin{equation}
\label{B3}
F_d = - \chi{ p' \over \rho }.
\end{equation}
where a prime denotes differentiation with respect to $z$.  Equations (\ref{B1}) and (\ref{B3}) then imply $F_d = \chi z$.  Any region with $\chi =$ const. is convectively unstable, since $F_c = 0$ implies that $\rho$ is constant while $p$ must decrease monotonically with height.  Moreover, any sharp increase of $\chi$ with height would exacerbate the instability, since it would imply a sheet of high density suspended in the atmosphere (according to eq.~[\ref{B2}]), i.e., a density inversion. (Strictly speaking, the upper part of the dense layer might be locally stable, if $\rho$ decreases steeply enough, but most of the atmosphere would be unstable.) Therefore, it seems reasonable to suppose that the model atmosphere is convective at all $z$.  If convection is efficient, then we can adopt the equation of state $p = A \rho^{4/3}$ and the structure of the atmosphere is independent of $\chi(z)$: 
\begin{equation}
\label{B4}
\rho = \rho_0 (1 - x^2)^3; \ \ \ p = p_0 (1 - x^2)^4 ,
\end{equation}
where $x = z/z_0$ and $z_0 = (8 A \rho_0^{1/3})^{1/2}$ is the height of the atmosphere.  The total column density of the atmosphere is
\begin{equation}
\label{B5}
\Sigma_0 = \rho_0 z_0 \int^1_0 (1-x^2)^3 dx = 0.457 \rho_0 z_0 .
\end{equation} 

Suppose that $\chi (z)$ satisfies the boundary conditions $\chi(0) = 1$ and $\chi(z_0) = \chi_0 \geq 1$.  Integrating eq.~(\ref{B2}) we have $F_c (z_0) = \Sigma_0 - \chi_0 z_0$.  But we also expect that all of the flux near the upper boundary should be carried by radiative diffusion, not convection, so that $F_c(z_0) = 0$.  This implies $\rho_0 = 2.19 \chi_0$.

Now, consider the fluxes carried by the atmosphere near the equator, i.e., at $x\ll 1$. Since $\chi \approx 1$ we have $F_d \approx z_0 x$ while $F_c \approx (\rho_0 - 1) z_0 x = (2.19\chi_0 - 1) z_0 x$.  Therefore the ratio of convective to diffusive flux near the equator is
\begin{equation}
\label{B6}
{F_c \over F_d} (x\ll 1) \approx 2.19 \chi_0 - 1 .
\end{equation}  
If $\chi_0$ represents the porosity associated with photon bubbles near the top of the disk, then $\chi_0 \sim \ell \gg 1$, where $\ell$ is the local Eddington enhancement factor.  The convective flux is therefore much larger than the diffusive flux, and is super-Eddington with respect to the {\it local} gravity at $z \ll z_0$.  This simple model shows that the development of photon bubbles near the photosphere can drive very strong convection near the equator.  Interestingly, even when photon bubbles are unimportant, $\chi_0 = 1$, the convective flux near the equator is slightly larger than the diffusive flux in this toy model, $F_c / F_d (x\ll 1) \approx 1.2$.

\end{document}